\documentclass[aps,prd,twocolumn,groupedaddress,superscriptaddress,longbibliography]{revtex4-1}
\usepackage{graphicx}
\usepackage{dcolumn}
\usepackage[dvipsnames]{xcolor}
\usepackage{amsmath}
\usepackage{physics}
\usepackage{amssymb}
\usepackage{appendix}
\usepackage{hyperref}
\newcommand{\mycomment}[1]{}
\usepackage{hhline}
\usepackage{longtable}
\usepackage{enumitem}
\usepackage{comment}

\begin{document}

\title{High-energy electron-positron beam collisions with large-angle disruptions}

\author{W. Zhang}
\email[]{wenlong.zhang@ecut.edu.cn}
\affiliation{Engineering Research Center of Nuclear Technology Application, Ministry of Education, East China University of Technology, Nanchang 330013, China}

\author{T. Grismayer}
\email[]{thomas.grismayer@tecnico.ulisboa.pt}

\author{L. O. Silva}
\email[]{luis.silva@tecnico.ulisboa.pt}
\affiliation{GoLP/Instituto de Plasmas e Fusão Nuclear, Instituto Superior Técnico, Universidade de Lisboa, Lisboa 1049-001, Portugal}

\date{\today}
\begin{abstract}
{We demonstrate that the beam and field dynamics in high-energy electron-positron collisions are governed by a new dimensionless parameter $\varepsilon$. In the high-disruption regime, particles are deflected transversely at angles of order $\varepsilon$ while simultaneously undergoing longitudinal deceleration which scales as $\varepsilon^2$, accompanied by the emergence of a longitudinal electric field with amplitude proportional to $\varepsilon$. We identify the large-angle disruptions ($\varepsilon \ge 1$) as a novel extreme regime in which transverse motion becomes strongly relativistic and the braking effect fully arrests, and subsequently reverses, beam propagation. Our theoretical model is in good agreement with three-dimensional electromagnetic particle-in-cell simulations. Existing beam-beam studies are valid only in the $\varepsilon \ll 1$ limit. For proposed plasma-based linear colliders operating in the quantum regime, notable beam-beam effects generate low-energy secondary particles whose $\varepsilon$ can approach unity; the collision dynamics and behavior of these particles may therefore require fully electromagnetic particle-in-cell simulations for accurate modeling.}
\end{abstract} 


\maketitle

\section{Introduction}
\label{sec: introduction}

In the pursuit of next-generation high-luminosity lepton colliders, the collective dynamics of electron-positron beam collisions at the Interaction Point (IP) represents a frontier where classical beam physics meets strong-field quantum electrodynamics (SF-QED) \cite{yokoya2005beam, Fabrizio2019, Zhang2023, Zhang2025}. The beam-beam interaction can serve as a rigorous testbed for beam and plasma instabilities, such as the kink instability \cite{yokoya2005beam, ZhangEPS2021, Zhang2025, Samsonov2021}, and provide a pathway to accessing the non-perturbative SF-QED regime \cite{Zhang2025, Yakimenko2019, Barklow2023}. Beyond theoretical exploration, high-luminosity and high-energy $e^-e^+$ collisions are the primary driver for next-generation collider designs that aim to address the grand challenges of particle physics \cite{Shiltsev2021, Gray2021, EurStrParPhys2022, Roser2023, Shiltsev2024, P5_Report, RevParPhys2026, 10TeV_design_initiative, Abramowicz2026, ALEGRO2024, ALEGRO2026}.

The collective beam motion, driven by the intense electromagnetic fields of the opposing beam, is the fundamental mechanism behind beam-beam effects at the IP. These effects include disruption that is characterized by a beam pinch in $e^-e^+$ collisions \cite{Chen1988} or repulsion in $e^-e^-$ collisions \cite{Zhang2025}. The beam fields also lead to beamstrahlung and electron-positron pair production \cite{Chen1989, Chen1992, yokoya2005beam, Esberg2014, Schulte2017, Samsonov2021, Tamburini2021, Song2021, Zhang2023, Barklow2023, Nguyen2024_LBNL_WarpX_GP, Zhang2025, He2025_SLAC_analytical_beamstrahlung, Madison_beambeam_2026}. The collision dynamics is currently known to be characterized by two dimensionless parameters: the disruption parameter $D$, which determines density pinch, field amplification, and luminosity enhancement \cite{Chen1988, Zhang2025, Mori1989}, and the quantum parameter ($\chi$ or $\Upsilon$) \cite{yokoya2005beam, Fabrizio2019, Zhang2023, Zhang2025, He2025_SLAC_analytical_beamstrahlung}, which dictates the strength of SF-QED processes \cite{Gonoskov2022, Fedotov2022, Zhang2020}.

Despite the depth of existing research, current theoretical models and widely used beam-beam codes \cite{Schulte1996, Chen1995_ABEL} rely on a critical simplification: the ``free-streaming" approximation. This assumes that particles maintain a constant longitudinal velocity at the speed of light $c$ throughout the collision, effectively treating transverse and longitudinal motions as fully separable. This assumption holds only when transverse deflections are negligible and the particles remain ultra-relativistic. 

In this paper, we demonstrate that the conventional framework for beam-beam studies fails as collisions approach a new regime of ``large-angle disruption." We show that the collision dynamics is governed by a third dimensionless parameter, introduced as $\varepsilon$ here. This parameter coincides with the particle's deflection angle and quantifies the coupling between transverse and longitudinal motion. While previous studies operate effectively in the limit $\varepsilon \ll 1$, we show that for $\varepsilon > 0.1$, transverse motion and current become comparable to or even dominate over their longitudinal counterparts.

During deflection, particles simultaneously undergo longitudinal deceleration (``braking effect"), accompanied by the emergence of a self-consistent longitudinal electric field. For $\varepsilon \ge 1$, this braking effect can completely arrest and further reverse the beam propagation. These dynamics not only increase the collision luminosity but also leave distinct imprints on the luminosity spectrum. Our findings are supported by a theoretical model that is in good agreement with 3-dimensional electromagnetic particle-in-cell (PIC) simulations.

The currently planned $e^-e^+$ linear colliders, like International Linear Collider (ILC) and Compact Linear Collider (CLIC), have typical $\varepsilon$ of $10^{-4}-10^{-3}$ (discussed in Sec. \ref{sec: discussion}). The SF-QED effects are generally weak. The conventional beam-beam codes like CAIN \cite{Yokoya1985, Chen1995_ABEL} and GUINEA-PIG \cite{Schulte1996}), designed for negligible deflections, remain fully valid for studying those machines. Regarding plasma-based colliders (Sec. \ref{sec: discussion}), they operate in the quantum regime. The notable beam-beam effects for round beams result in beamstrahlung and the generation of low-energy particles whose $\varepsilon$ can approach unity. The detailed collision dynamics and exact behaviors of produced secondary particles in such colliders may need to be studied using self-consistent PIC simulations.

\section{Theoretical model}
\label{sec: model}

\begin{figure}
\includegraphics[width=8cm,height=5.3cm]{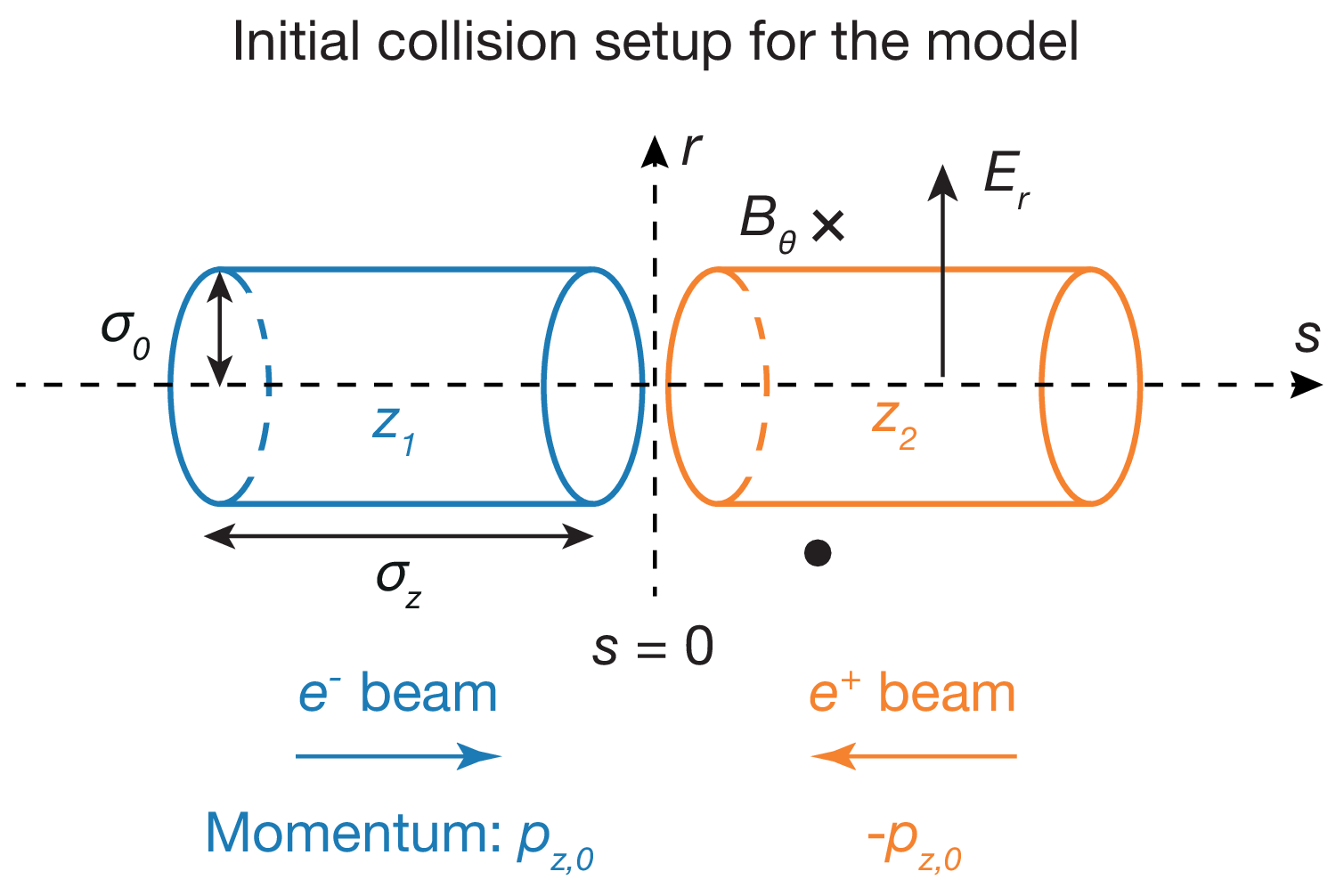}
\caption{(Color online). Initial setup (at $t=0$) for a typical collision considered by our theoretical model. The beams have cylindrical, uniform-density profiles.}
\label{fig: schematic_collision_setup_model}
\end{figure}

Our model considers an electron-positron collision between two round, cylindrical beams of uniform density. A schematic plot is provided in Fig. \ref{fig: schematic_collision_setup_model}. This idealized configuration allows for a transparent analytical treatment \cite{Zhang2025}, the results of which can be mapped to realistic Gaussian-beam profiles via established profile transformations \cite{Zhang2023, Zhang2025}.

The interaction is described in a cylindrical coordinate system $(r, \theta, s)$, where $s$ denotes the longitudinal axis in the center-of-mass frame \cite{Zhang2025}. We define $z_j$ ($j=1, 2$) as the longitudinal coordinate co-moving with the electron and positron beams, respectively. Under this convention, each beam occupies the interval $-\sigma_z \le z_j \le 0$, where $z_j = 0$ represents the beam front and $z_j = -\sigma_z$ the tail.

The collision between an electron slice at $z_1$ and a positron slice at $z_2$ occurs at a specific time $t$ and location $s$, governed by the kinematic relations $s = z_1 + ct = -z_2 - ct$ \cite{Zhang2025, Chen1988}. Each beam is characterized by its initial energy $\mathcal{E}_0$, transverse radius $\sigma_0$, longitudinal length $\sigma_z$, and particle number $N_0$, corresponding to a uniform density $n_0 = N_0 / (\pi \sigma_0^2 \sigma_z)$. For Gaussian-beam collisions, $\sigma_0$ and $\sigma_z$ denote the root mean square (RMS) width and length of the beams. 

\subsection{Particle motion in transverse fields}
\label{subsec: particle_motion_in_collision}

The beams are considered to be mono-energetic, with Lorentz factor $\gamma_0 = \mathcal{E}_0/mc^2=1/\sqrt{1-\beta_{z,0}^2} \gg 1$, where $m$ is the mass of an electron and $\beta_{z,0}=v_{z,0}/c$ with $v_{z,0}$ the initial streaming velocity. $p_{z,0} = \gamma_0mc\beta_{z,0}$ is the momentum. The undisrupted self-fields of a beam are given by \cite{Zhang2025}
\begin{equation}
E_{r,0}(r)=\pm E_0\frac{r}{\sigma_{0}}, \ B_{\theta,0}(r)=\beta_{z,0} E_{r,0}(r) , \ E_{z,0} = 0,
\label{Eq: undisrupted_selffields}
\end{equation}
where the signs $\pm 1$ are for positron and electron beams, respectively, and $E_0=2\pi e n_0\sigma_{0}$ is the maximum electric field, with $e$ the electron charge.  

Without beamstrahlung (i.e., photon emission \cite{yokoya2005beam, Zhang2023}), the equation of motion for an electron under the undisrupted self-fields of the opposing positron beam reads
\begin{subequations}
\begin{align}
     & \frac{dp_r}{dt} = -e(1-\beta_z \beta_z^+) E_0\frac{r}{\sigma_{0}}, \label{Eq: equation_pr} \\
     & \frac{dp_z}{dt} = -e\beta_r \beta_z^+ E_0\frac{r}{\sigma_{0}}, \label{Eq: equation_pz} \\
     & \frac{d\gamma}{dt} = - \frac{eE_0}{mc}\beta_r\frac{r}{\sigma_0},
     \label{Eq: equation_gamma}
\end{align}
\label{Eq: equation_motion_p_gamma}
\end{subequations}
with $\beta_z^+=v_z^+/c$, where $v_z^+$ is the velocity of the positron beam; $\beta_z = v_z/c$ and $\beta_r=v_r/c=p_r/\gamma mc$. Equation \eqref{Eq: equation_motion_p_gamma} can also be written as
\begin{subequations}
\begin{align}
\frac{d\beta_r}{dt}  & = -\frac{1}{2}(1 - \beta_z\beta_z^+ - \beta_r^2)\frac{\gamma_0}{\gamma}\frac{r}{\sigma_0}\frac{\varepsilon}{\tau_D}, \label{Eq: equation_beta_r} \\
\frac{d\beta_z}{dt} & = \frac{1}{2}(\beta_z - \beta_z^+)\beta_r \frac{\gamma_0}{\gamma}\frac{r}{\sigma_0}\frac{\varepsilon}{\tau_D} , \label{Eq: equation_beta_z} \\
\frac{1}{\gamma_0}\frac{d\gamma}{dt} & = -\frac{1}{2}\beta_r\frac{r}{\sigma_0} \frac{\varepsilon}{\tau_D}, \label{Eq: equation_gamma_transformed}
\end{align}
\label{Eq: equation_motion_beta_gamma}
\end{subequations}
where $\varepsilon$ is defined in Eq. \eqref{Eq: eps_definition} and $\tau_D$ is the characteristic time scale of beam disruption given by
\begin{equation}
    \tau_D = \frac{\sqrt{\gamma_0}} {\omega_b},
    \label{Eq: tau_D_definition}
\end{equation}
with $\omega_b=\sqrt{4\pi e^2n_0/m}$ the beam plasma frequency. The disruption parameter $D$ \cite{Zhang2025} can be defined as the ratio between the nominal collision time $\tau_{col} = \sigma_{z}/c$ and the disruption time $\tau_D$, i.e.,
\begin{equation}
D=\frac{\tau_{col}^2}{\tau_D^2} = \left(\frac{\sigma_z}{\sqrt{\gamma_0}c/\omega_b} \right)^2.
\label{Eq: D_definition}
\end{equation}
Equation \eqref{Eq: equation_beta_z} reveals that $d\beta_z/dt < 0$, demonstrating that particles necessarily undergo longitudinal deceleration as they are deflected. This ``braking effect" signifies a fundamental coupling between the transverse and longitudinal dynamics. According to Eq. \eqref{Eq: equation_motion_beta_gamma}, the particle motion is effectively governed by $\varepsilon$, which is the new dimensionless parameter introduced in this study, i.e.,
\begin{equation}
\varepsilon = \frac{\sigma_0}{\sqrt{\gamma_0}c/\omega_b} = \frac{\sigma_0}{c\tau_D} = \sqrt{D}\frac{\sigma_{0}}{\sigma_z}.
\label{Eq: eps_definition}
\end{equation}
The definition of $\varepsilon$ [Eq. \eqref{Eq: eps_definition}] is valid for any disruption ($D$), since the equation of motion [Eq. \eqref{Eq: equation_motion_beta_gamma}] holds at any $D$ for $t\le \tau_D$ (before significant beam disruption occurs). We note that $\tau_{col}<\tau_D$ for $D<1$. Our study here is primarily focused on the disruptive regime with $D\ge 1$.

While the disruption parameter $D$ characterizes the number of oscillations along the longitudinal direction [Eq. \eqref{Eq: D_definition}], $\varepsilon$ evaluates the transverse beam dimension $\sigma_0$ normalized by the relativistically-corrected skin depth of the beam, defined as $\sqrt{\gamma_0}c/\omega_b$. Equation \eqref{Eq: eps_definition} indicates that $\varepsilon$ is also the ratio between transverse size $\sigma_0$ and characteristic longitudinal focal length ($c\tau_D$ or $\sigma_z/\sqrt{D}$). To facilitate practical application, Eq. \eqref{Eq: eps_definition} can be recast into the following engineering form:
\begin{equation}
\varepsilon = \eta \sqrt{\frac{N_0[10^{10}]}{\mathcal{E}_0[\mathrm{GeV}] \, \sigma_z[\mathrm{\mu m}]}},
\label{Eq: eps_engineering_definition}
\end{equation}
where $\eta$ is a geometric factor that depends on the beam profile, such that $\eta = 0.24$ for uniform beams considered in our model. For Gaussian beams, $\eta = 0.15$, where a profile transformation is required \cite{Zhang2025}.

In the limit $\varepsilon \ll 1$, Eq. \eqref{Eq: equation_motion_beta_gamma} implies that velocity changes at $t \simeq \tau_D$ satisfy $|\Delta \beta_r| \ll 1$ and $|\Delta \beta_z| \ll 1$. Consequently, the longitudinal deceleration is negligible, and the transverse deflection remains a minor perturbation to the longitudinal streaming. Under these conditions, the particle energy remains nearly constant ($|\Delta \gamma| \ll \gamma_0$). This regime forms the physical basis for the ``free-streaming" approximation ($v_z \equiv \pm c$) and the assumption of constant energy utilized in both previous theoretical models \cite{Chen1988, yokoya2005beam, Zhang2025} and standard numerical beam-beam codes \cite{Schulte1996,Yokoya1985}.

In contrast, when $\varepsilon$ is not negligible (approaching unity), the particle dynamics enters a nonlinear regime where transverse and longitudinal motions become strongly coupled. This coupling breaks the free-streaming approximation, giving rise to the drastic beam-field features that are the focus of this study.

\subsection{Harmonic oscillator for small-angle disruptions}
\label{subsec: harmonic_oscillator}
For $\varepsilon\ll 1$, one can assume $\gamma \simeq \gamma_0$, $\beta_z \simeq 1$, $\beta_z^+ \simeq -1$, and $|\beta_r| \ll \beta_z$ as analyzed in Sec. \ref{subsec: particle_motion_in_collision}. Equation \eqref{Eq: equation_beta_r} reduces to  
\begin{equation}
    \frac{d^2r}{dt^2} = -\frac{1}{\tau_D^2}r,
    \label{Eq: diff2_r_t}
\end{equation}
which is the equation of an ideal harmonic oscillator. The solution for the position and the velocity are
\begin{equation}
    r = r_0\cos \left( \frac{t}{\tau_D} \right), \ \beta_r = \frac{v_r}{c} = -\frac{r_0}{c\tau_D}\sin \left(\frac{t}{\tau_D}\right),
    \label{Eq: harmonic_oscillator_solution}
\end{equation}
where $r_0$ is the initial position. 

Our study is focused on the collision dynamics in the disruptive regime where $D\ge 1$. In this condition, a particle will approach the axis at $t\simeq \tau_D$. The maximum deflection velocity ($\beta_{r,\, \mathrm{max}}$) and the disruption angle ($\theta_D$) are
\begin{equation}
 \beta_{r,\, \mathrm{max}} = -\frac{\sigma_0}{c\tau_D} = -\varepsilon, \  \theta_D = \frac{|\beta_{r,\, \mathrm{max}}|}{\beta_z} =  \varepsilon.
 \label{Eq: deflection_angle}
\end{equation}  
Consequently, $\varepsilon$ physically represents both the deflection velocity and the disruption angle. We note that $\theta_D\sim D\sigma_0/\sigma_z=\sqrt{D}\varepsilon$ for weak disruptions ($D<1$) \cite{Schulte1996, Zhang2023}.

In the $\varepsilon \ll 1$ regime, Eq. \eqref{Eq: deflection_angle} yields $|\beta_r| \ll 1$, substantiating the assumption (for Eq. \eqref{Eq: diff2_r_t}) that transverse deflection is negligible ($|\beta_r| \ll \beta_z$). Conversely, for $\varepsilon > 0.1$, the oscillatory motion becomes relativistic ($v_r \sim -c$) and comparable to the longitudinal streaming.

The linearized harmonic motion described here has served as the foundation for various beam-related studies, including derivations of density evolution during collisions (see Sec. \ref{subsec: pinch_density_currents}) and the analysis of collective instabilities and beam movements, such as kink, hosing, and betatron oscillations \cite{Rosenzweig1989, yokoya2005beam, Moreira2023, Bilbao2025}.

\subsection{Analogy to electron motion in an electromagnetic wave}

\label{subsec: analogy_to_electron_motion_EM_wave}
When an electron interacts with an electromagnetic wave of frequency $\omega$ and amplitude $E$, the electron oscillates with typical momentum $p$ given by \cite{Gibbon2005}
\begin{equation}
    \frac{p}{mc} = a_0 = \frac{eE}{m\omega c},
    \label{Eq: vos_a0_laser}
\end{equation}
where the normalized amplitude $a_0$ characterizes the non-linearity of the motion. 

In the context of beam collisions, these parallels lead naturally to the question of whether an equivalent normalized vector potential can be defined for the beam fields. A previous approach \cite{Fabrizio2019} proposed such a definition by using $c/\sigma_z$ as the characteristic frequency, effectively treating the beam fields as a mono-cycle pulse.

In this study, we define the equivalent normalized amplitude by utilizing the characteristic frequency of the harmonic oscillator (formulated in Sec. \ref{subsec: harmonic_oscillator}), $\omega_D = (2\tau_D)^{-1}$:
\begin{equation}
    a_{beam} = \frac{eE_0}{m\omega_D c} = \gamma_0\varepsilon.
    \label{Eq: a0_beam_beam}
\end{equation}
In this way, the typical radial momentum becomes $p_r/mc = a_{beam}$

By manipulating the equations of motion [Eqs. \eqref{Eq: equation_motion_p_gamma} and \eqref{Eq: equation_motion_beta_gamma}], we derive two fundamental conservation relations:
\begin{subequations}
\begin{align}
\frac{d}{dt}(\gamma + p_z/mc) & = 0, \label{Eq: diff_gamma_pz} \\
\frac{d}{dt}(\beta_r^2/2 + \beta_z) & = 0. \label{Eq: diff_betar2_betaz}
\end{align}
\label{Eq: diff_eq_for_invariants}
\end{subequations}
These relations are obtained under the assumptions of $|\beta_r| \ll \beta_z$, $\gamma \simeq \gamma_0$, and $\beta_z^+ \simeq -1$. As discussed in Sec. \ref{subsec: harmonic_oscillator}, these approximations are fully consistent with the $\varepsilon \ll 1$ regime.

Equation \eqref{Eq: diff_eq_for_invariants} identifies two key invariants of the particle motion:
\begin{subequations}
\begin{align}
\gamma + \frac{p_z}{mc} & \equiv 2\gamma_0, \label{Eq: invariant_gamma_pz} \\
\frac{1}{2}\beta_r^2 + \beta_z & \equiv \beta_{z,0} \simeq 1. \label{Eq: invariant_betar_betaz}
\end{align}
\label{Eq: two_invariants}
\end{subequations}
Remarkably, these invariants are identical to those found for a charged particle interacting with an electromagnetic plane wave \cite{Gibbon2005}. This identity further solidifies the analogy between disruption-induced particle motion in a collision and high-intensity laser-matter interaction.

The invariant in Eq. \eqref{Eq: invariant_gamma_pz} reveals that the particle gains kinetic energy at the direct expense of its longitudinal momentum. This energy-momentum coupling can be expressed as:
\begin{equation}
    \Delta \gamma = \gamma - \gamma_0 = \gamma_0 - \frac{p_z}{mc} = -\frac{\Delta p_z}{mc}.
    \label{Eq: Delta_gamma_Delta_pz}
\end{equation}
Similarly, Eq. \eqref{Eq: invariant_betar_betaz} demonstrates that transverse acceleration is fueled by longitudinal deceleration. By substituting the peak deflection velocity $\beta_r \simeq -\varepsilon$ [Eq. \eqref{Eq: deflection_angle}], the deceleration is found to scale as:
\begin{equation}
    \Delta \beta_z = \beta_z - \beta_{z,0} = -\frac{1}{2}\beta_r^2 \simeq -\frac{1}{2}\varepsilon^2.
    \label{Eq: approximate_Delta_beta_z}
\end{equation}

By combining Eqs. \eqref{Eq: Delta_gamma_Delta_pz} and \eqref{Eq: approximate_Delta_beta_z}, and utilizing the relation $\Delta p_z \simeq (\Delta \gamma \beta_{z,0} + \gamma_0 \Delta \beta_z)mc$, we can derive the overall scalings for the energy and momentum shifts:
\begin{subequations}
\begin{align}
\Delta \gamma & \simeq \frac{1}{4}\varepsilon ^2\gamma_0, \label{Eq: approximate_Delta_gamma} \\
\Delta p_z & \simeq -\frac{1}{4}\varepsilon ^2 p_{z,0}. \label{Eq: approximate_Delta_pz}
\end{align}
\label{Eq: approximate_Deltagamma_Deltapz}
\end{subequations}
These semi-analytical scalings establish that the braking effect and the resulting energy redistribution are second-order in $\varepsilon$. Remarkably, these scaling results are in precise agreement with the formal perturbation solution derived in the next section.

\subsection{Perturbative solution for finite $\varepsilon$}
\label{subsec: perturbation_solution}

While the linear harmonic oscillator in Sec. \ref{subsec: harmonic_oscillator} is a reduced model strictly valid for $\varepsilon \ll 1$, particles in the finite-$\varepsilon$ regime behave as modified, nonlinear oscillators. To capture this behavior, we solve the equation of motion [Eq. \eqref{Eq: equation_motion_beta_gamma}] for finite $\varepsilon$ (while maintaining $\varepsilon < 1$) using a perturbation approach. Here, we isolate the effects of the Lorentz force $F_\mathrm{L}$ and do not yet consider the induced longitudinal field $E_z$. The resulting perturbation solutions are:
\begin{subequations}
\begin{align}
\Delta \gamma ^{F_\mathrm{L}} & = \frac{1}{4}\varepsilon^2\frac{r_0^2}{\sigma_0^2}\left(1-\cos^2 (t/\tau_D) \right)\gamma_0, \label{Eq: perturbation_solution_Delta_gamma} \\
\Delta p_z ^{F_\mathrm{L}} & = -\frac{1}{4}\varepsilon^2\frac{r_0^2}{\sigma_0^2}\left(1-\cos^2 (t/\tau_D)\right)p_{z,0}, \label{Eq: perturbation_solution_Delta_pz} \\
\Delta \beta_z ^{F_\mathrm{L}} & = -\frac{1}{2}\varepsilon^2\frac{r_0^2}{\sigma_0^2}\left(1-\cos^2 (t/\tau_D)\right)\beta_{z,0}, \label{Eq: perturbation_solution_Delta_betaz}
\end{align}
\label{Eq: perturbation_solution_Delta}
\end{subequations}
where $\Delta \gamma ^{F_\mathrm{L}}$, $\Delta p_z ^{F_\mathrm{L}}$, and $\Delta \beta_z ^{F_\mathrm{L}}$ denote energy gain, longitudinal momentum loss, and longitudinal deceleration, respectively. The complete derivation of Eq. \eqref{Eq: perturbation_solution_Delta} is provided in Appendix \ref{SM_sec: perturbation_solution}.

Equation \eqref{Eq: perturbation_solution_Delta} confirms that the perturbations to the particle motion scale as $\varepsilon^2$, providing a formal derivation for the semi-analytical scalings achieved in Sec. \ref{subsec: analogy_to_electron_motion_EM_wave} [cf. Eqs. \eqref{Eq: approximate_Delta_beta_z} and \eqref{Eq: approximate_Deltagamma_Deltapz}]. Furthermore, these solutions reveal a spatial dependence on the initial radial position $r_0$, with the perturbations growing as $(t/\tau_D)^2$ in the early interaction phase ($t \ll \tau_D$).

To illustrate the spatial non-uniformity of these effects, we evaluate the relative momentum and energy shifts using Eq. \eqref{Eq: perturbation_solution_Delta}, for on-axis ($r_0=0$) and peripheral ($r_0=\sigma_0$) particles at the characteristic disruption time $t=\tau_D$:
\begin{subequations}
\begin{align} 
\left.\frac{\Delta p_z ^{F_\mathrm{L}}}{p_{z,0}} \right\vert_{r_0=0} & = 0, \ \left.\frac{\Delta p_z ^{F_\mathrm{L}}}{p_{z,0}} \right\vert_{r_0=\sigma_0} = -0.18\varepsilon^2, \label{Eq: Delta_pz_without_Ez} \\
\left.\frac{\Delta \gamma ^{F_\mathrm{L}}}{\gamma_0} \right\vert_{r_0=0} & = 0, \ \left.\frac{\Delta \gamma ^{F_\mathrm{L}}}{\gamma_0} \right\vert_{r_0=\sigma_0} = 0.18\varepsilon^2. \label{Eq: Delta_gamma_without_Ez}
\end{align}
\label{Eq: Delta_pz_gamma_without_Ez_diff_r0}
\end{subequations}
These results indicate that, under the action of Lorentz force alone, the particles close to the aixs remain largely unaffected, while peripheral particles suffer a loss of momentum $p_z$ but gain energy $\gamma$. However, as we will demonstrate in Sec. \ref{subsec: field_Ez}, the emergence of longitudinal field $E_z$ alters this picture, imposing a substantial impact on the dynamics of both the individual particles and the collective beam.

\subsection{Beam density and current}
\label{subsec: pinch_density_currents}

The system density and current at time $t$ and location $s$ are determined by the joint contribution from the local electron and positron slices which are colliding. Since we focus on the small-$\varepsilon$ regime ($\varepsilon<1$) here, the braking effect of longitudinal deceleration is weak (cf. Eqs. \eqref{Eq: approximate_Delta_beta_z} and \eqref{Eq: perturbation_solution_Delta_betaz}) and thus is neglected for a simplified analysis. See the beginning of Sec. \ref{sec: model} for a detailed description of the coordinate convention and collision configuration for our model. Based on the oscillator model [Eq. \eqref{Eq: harmonic_oscillator_solution}], the radial position of a beam element evolves as
\begin{equation}
    r_j(s, t)=r_{0,j}\cos \left(\frac{\Delta t_j}{\tau_D}\right) \simeq r_{0,j}\left[1-\frac{1}{2}\left(\frac{\Delta t_j}{\tau_D} \right)^2 \right],
    \label{Eq: r_in_a_collision}
\end{equation}
where $\Delta t_j$ represents the time elapsed since the particle (from $z_j$) crossed the front of the opposing beam at the moment $t_{0,j}$. $\Delta t_j$ and $t_{0,j}$ are given by
\begin{equation}
    t_{0,j} = -\frac{z_j}{2c}, \ \Delta t_j = t - t_{0,j} = \frac{1}{2}\left( t \pm \frac{s}{c} \right).
    \label{Eq: Delta_tj}
\end{equation}
During the collision, the electron and positron beams occupy the regions of $(ct-\sigma_z)\le s \le ct$ and $-ct \le s \le (\sigma_z-ct)$, respectively. 

As the particles are deflected inward according to Eq. \eqref{Eq: r_in_a_collision}, the beams undergo a transverse pinch. The resulting density evolution is governed by particle conservation \cite{Zhang2025}, $n_j r_j dr_j = n_0 r_{0,j} dr_{0,j}$, which yields:
\begin{equation}
    n_j(s,t) = \frac{n_0}{\left[1-\frac{1}{2}\left(\frac{\Delta t_j}{\tau_D} \right)^2 \right]^2} \simeq n_0 \left[1+\left(\frac{\Delta t_j}{\tau_D} \right)^2 \right] .
    \label{Eq: beam_density_pinch}
\end{equation}

The beam pinch further perturbs the electric currents, which in turn drive the fields discussed in Sec. \ref{subsec: field_Ez}. We first define the $0$th-order currents, longitudinal ($j_{z0}$) and transverse ($j_{r0}$), for the idealized case assuming that the disruption does not occur:
\begin{subequations}
\begin{align}
& j_{z0} = j_{z,1} + j_{z,2} = -2j_0, \label{Eq: undisrupted_jz} \\
& j_{r0} = j_{r,1} + j_{r,2} = 0, \label{Eq: undisrupted_jr}
\end{align}
\label{Eq: undisrupted_j}
\end{subequations}
where $j_{z,1} = j_{z,2} = -j_0$ with $j_0 = n_0ec$, and $j_{r,1} = j_{r,2} = 0$.

The perturbation to the longitudinal current due to disruption, $\delta j_z$, is derived from the density evolution given by Eq. \eqref{Eq: beam_density_pinch}:
\begin{eqnarray}
\delta j_z(s,t) & = & -n_1ec - n_2ec - j_{z0} \nonumber \\
& \simeq & -j_0\frac{1}{2}D\left( \frac{t^2}{\tau_{col}^2} + \frac{s^2}{\sigma_z^2}\right).
\label{Eq: perturbation_jz}
\end{eqnarray}
Equation \eqref{Eq: perturbation_jz} reveals that the magnitude of $\delta j_z$ scales linearly with $D$, as the disruption parameter directly governs the strength of the beam pinch.

Simultaneously, the beam pinch imparts a transverse velocity to the beam elements. Differentiating Eq. \eqref{Eq: r_in_a_collision} with respect to time, the radial velocity for a beam element is:
\begin{eqnarray}
v_{r,j}(s,t) = \frac{dr_j(s,t)}{dt} & \simeq & -r_{0,j}\frac{\Delta t_j}{\tau_\textrm{\tiny D}}\frac{d}{dt}\left(\frac{\Delta t_j}{\tau_\textrm{\tiny D}}\right) \nonumber \\
& = & -\frac{1}{2}\frac{r_{0,j}}{\tau_\textrm{\tiny D}}\frac{\Delta t_j}{\tau_\textrm{\tiny D}} \nonumber \\
& = & -\frac{1}{2}\varepsilon \frac{r_{0,j}}{\sigma_0}\frac{\Delta t_j}{\tau_\textrm{\tiny D}}c,
\label{Eq: vr_beam_element}
\end{eqnarray}
where $\Delta t_j$ is given in Eq \eqref{Eq: Delta_tj}. Note that Eq. \eqref{Eq: vr_beam_element} does not contradict the velocity scaling obtained in Eq. \eqref{Eq: harmonic_oscillator_solution} (and Eq. \eqref{Eq: deflection_angle}). Equation \eqref{Eq: vr_beam_element} solves velocity of the corresponding beam element which is flowing through a location $s$ at the moment $t$, and Eq. \eqref{Eq: harmonic_oscillator_solution} gives velocity of a particle following its trajectory.

The transverse motion drives a net radial current, $\delta j_r$:
\begin{equation}
\delta j_r (r,s) = -n_1 ev_{r,1} + n_2ev_{r,2} \simeq j_0 \sqrt{D}\varepsilon \frac{r}{\sigma_0}\frac{s}{\sigma_z}.
\label{Eq: perturbation_jr}
\end{equation}
Equation \eqref{Eq: perturbation_jr} shows that $\delta j_r \propto \sqrt{D}\varepsilon rs$, with a linear dependence on radial position $r$ and longitudinal location $s$. Crucially, our result identifies $\varepsilon$ as a dominant factor in dictating the transverse current, since $\delta j_r\propto \varepsilon$. In our derivation for $\delta j_r$ [Eq. \eqref{Eq: perturbation_jr}], $\delta j_r$ is found to possess a weak time-dependence which is neglected in Eq. \eqref{Eq: perturbation_jr}. The above features of $\delta j_r$ are supported by our particle-in-cell simulations.

It is worth comparing the radial and longitudinal currents, since they dictate the self-consistent field evolution (see Sec. \ref{subsec: field_Ez}). The amplitudes of the current perturbations at $t=\tau_D$ can be obtained using Eqs. \eqref{Eq: perturbation_jz} and \eqref{Eq: perturbation_jr}:
\begin{equation}
    (\delta j_z)_{\mathrm{max}} = -j_0, \ (\delta j_r)_{\mathrm{max}} = \pm \varepsilon j_0,
    \label{Eq: max_jz_jr}
\end{equation}
which yields the simple relation
\begin{equation}
    \left\vert \frac{(\delta j_r)_{\mathrm{max}}}{(\delta j_z)_{\mathrm{max}}}\right\vert = \varepsilon.
    \label{Eq: ratio_delta_jr_jz}
\end{equation}
Consequently, $\varepsilon$ directly dictates the relative amplitudes of radial and longitudinal current perturbations. We can further investigate the total system currents which are given by \begin{subequations}
\begin{align}
& j_z = j_{z0} + \delta j_z, \label{Eq: tot_jz} \\
& j_r = j_{r0} + \delta j_r. \label{Eq: tot_jr}
\end{align}
\label{Eq: tot_j}
\end{subequations}
Using Eq. \eqref{Eq: undisrupted_j} and Eq. \eqref{Eq: max_jz_jr}, the peak amplitudes of the total currents follow a similar relation
\begin{equation}
    \left\vert \frac{(j_r)_{\mathrm{max}}}{(j_z)_{\mathrm{max}}}\right\vert = \frac{1}{3}\varepsilon .
    \label{Eq: ratio_total_jr_jz}
\end{equation}
Equation \eqref{Eq: ratio_total_jr_jz} demonstrates that the radial current becomes comparable to the longitudinal current for $\varepsilon > 0.3$.

\subsection{Emergence of $E_z$}
\label{subsec: field_Ez}

The self-consistent evolution of the electromagnetic fields is governed by Maxwell's equations. In the cylindrical geometry of a beam-beam collision, the relevant field components evolve as:
\begin{subequations}
    \begin{align}
    \frac{\partial E_r}{\partial t} & = -c \frac{\partial B_\theta}{\partial s} -4\pi j_r, \label{Eq: equation_Er}\\
    \frac{\partial E_z}{\partial t} & = c\frac{1}{r}\frac{\partial }{\partial r}(rB_\theta) - 4\pi j_z, \label{Eq: equation_Ez}\\ 
    \frac{\partial B_\theta}{\partial t} & = c\left(\frac{\partial E_z}{\partial r} - \frac{\partial E_r}{\partial s} \right),
    \label{Eq: equation_Btheta}   
    \end{align}
    \label{Eq: Maxwell_equations}
\end{subequations}
while the remaining components ($B_z$, $B_r$, and $E_\theta$) vanish due to symmetry. The total system fields result from superimposing the fields of the two colliding beams. Analogous to the current decomposition, the total fields are expressed as the sum of the undisrupted $0$th-order terms and the disruption-induced perturbations:
\begin{subequations}
    \begin{align}
    E_r & = E_{r0}+\delta E_r, \label{Eq: Er_decompose}\\
    B_\theta & = B_{\theta 0} + \delta B_\theta, \label{Eq: Btheta_decompose}\\ 
    E_z & = E_{z 0} + \delta E_z. \label{Eq: Ez_decompose}   
    \end{align}
    \label{Eq: fields_decompose}
\end{subequations}
In the interaction region, the $0$th-order terms of the system fields are jointly determined by the undisrupted self-fields of the two beams (given by Eq. \eqref{Eq: undisrupted_selffields}), i.e.,
\begin{subequations}
    \begin{align}
    E_{r0} & = \sum_{j=1,2} E_{r,0}^j = 0, \label{Eq: Er0_total_fields} \\
    B_{\theta 0} & = \sum_{j=1,2} B_{\theta,0}^j = -2E_0r/\sigma_0, \label{Eq: Btheta0_total_fields} \\ 
    E_{z0} & = \sum_{j=1,2} E_{z,0}^j = 0. \label{Eq: Ez0_total_fields}   
    \end{align}
    \label{Eq: undisrupted_total_fields}
\end{subequations}

The self-consistent field perturbations arise from a coupled evolution: the transverse current $\delta j_r$ initially drives $\delta E_r$ [Eq. \eqref{Eq: equation_Er}], which subsequently induces $\delta B_\theta$ via Eq. \eqref{Eq: equation_Btheta}. Our analysis indicates that $\delta E_r$ and $\delta B_\theta$ eventually reach comparable magnitudes. Notably, the growth of $\delta E_r$ is moderated by $\delta B_\theta$, which partially cancels the contribution of $\delta j_r$ in Eq. \eqref{Eq: equation_Er}. By accounting for this partial cancellation—effectively reducing the $\delta j_r$ contribution by half—we derive the perturbed transverse electric field as:
\begin{eqnarray}
\delta E_r(r,s,t) & \simeq & \frac{1}{2} \int_{|s|/c}^{t}(-4\pi \delta j_r) dt' \nonumber \\
& \simeq & -E_0D\frac{r}{\sigma_0} \frac{s}{\sigma_z}\frac{t}{\tau_{col}}.
\label{Eq: delta_Er}
\end{eqnarray}
Equation \eqref{Eq: delta_Er} shows a good agreement with our particle-in-cell simulation results, particularly for the elongated beam geometries ($\sigma_0 \ll \sigma_z$) typical of collider designs.

As we demonstrate in the following sections, the induced longitudinal field is largely insensitive to $s$ (i.e., $\partial \delta E_z/\partial s \simeq 0$) for the high-aspect-ratio beams ($\sigma_0 \ll \sigma_z$) considered here. Under this condition, the divergence of the electric field can be evaluated using Eq. \eqref{Eq: delta_Er}, yielding the relation:
\begin{equation}
    \nabla \cdot E \simeq \frac{1}{r}\frac{\partial (r\delta E_r)}{\partial r} = -2D\frac{E_0}{\sigma_0}\frac{s}{\sigma_z}\frac{t}{\tau_{col}} = 4\pi e(n_2-n_1),
    \label{Eq: Gauss_law_satisfaction}
\end{equation}
where the density $n_1$ and $n_2$ are given by Eq. \eqref{Eq: beam_density_pinch}. Equation \eqref{Eq: Gauss_law_satisfaction} demonstrates that the analytical solution for $\delta E_r$ [Eq. \eqref{Eq: delta_Er}] satisfies Gauss’s Law. 

While the transverse field $\delta E_r$ satisfies Gauss’s Law, a mismatch between $\delta B_\theta$ and $\delta j_z$ induces a longitudinal electric field $\delta E_z$ via Eq. \eqref{Eq: equation_Ez}. Our particle-in-cell simulations indicate that $\delta E_z$ is maximized at the collision center ($r=0$, $s=0$). For high-aspect-ratio beams ($\sigma_0 \ll \sigma_z$), the field $\delta E_z$ develops a characteristic parabolic radial profile, which obeys
\begin{equation}
    \delta E_z = \delta E_{z0}(t) \left( 1-\frac{r^2}{\sigma_0^2}\right).
    \label{Eq: full_delta_Ez}
\end{equation}
In the paraxial region ($r \ll \sigma_0$), the coupling between Eqs. \eqref{Eq: equation_Ez} and \eqref{Eq: equation_Btheta} leads to a second-order differential equation for the axial field amplitude:
\begin{equation}
    \frac{\partial ^2\delta E_{z0}(t)}{\partial (t/\tau_{col})^2} = - 4\frac{\sigma_z^2}{\sigma_0^2}\delta E_{z0}(t) + 4E_0D\frac{\sigma_z}{\sigma_0}\frac{t}{\tau_{col}} .
    \label{Eq: diff2_delta_Ez0} 
\end{equation}
The solution to Eq. \eqref{Eq: diff2_delta_Ez0} reads
\begin{equation}
    \delta E_{z0}(t) = E_0\left[\sqrt{D}\varepsilon \frac{t}{\tau_{col}} - \frac{1}{2}\varepsilon ^2 \sin \left( 2\frac{\sigma_z}{\sigma_0}\frac{t}{\tau_{col}}\right) \right].
    \label{Eq: delta_Ez0}
\end{equation}

The field $\delta E_z$ plays a critical role in the collision dynamics by inducing losses in both longitudinal momentum and particle energy. These $E_z$-caused shifts are given by:
\begin{align}
 \Delta p_z ^{E_z} & \simeq  -\frac{1}{4}\varepsilon^2 \Bigg[\left(1-\frac{1}{2}\frac{r_0^2}{\sigma_0^2}\right)\frac{t^2}{\tau_D^2} \label{Eq: Delta_pz_Ez} \\
 & \ \ \ \ + \frac{1}{4}\frac{r_0^2}{\sigma_0^2} \left(1 - \cos\frac{2t}{\tau_D} - \frac{2t}{\tau_D}\sin \frac{2t}{\tau_D}\right) \Bigg]p_{z,0}, \nonumber
\end{align}
and
\begin{equation}
    \Delta \gamma ^{E_z} \simeq  \frac{\Delta p_z ^{E_z}}{p_{z,0}}\gamma_0.
    \label{Eq: Delta_gamma_Ez}
\end{equation}
The derivations for Eqs. \eqref{Eq: Delta_pz_Ez} and \eqref{Eq: Delta_gamma_Ez} are provided in Appendix \ref{SM_sec: Deltapz_Deltagamma_by_Ez}.

To highlight the spatial distribution of these losses caused by $E_z$, we examine the on-axis ($r_0=0$) and peripheral ($r_0=\sigma_0$) particles at $t=\tau_D$ and get:
\begin{subequations}
    \begin{align}
     \left.\frac{\Delta p_z ^{E_z}}{p_{z,0}}\right\vert_{r_0=0} & = -\frac{1}{4}\varepsilon^2, \ \left.\frac{\Delta p_z ^{E_z}}{p_{z,0}}\right\vert_{r_0=\sigma_0} & = -0.1\varepsilon^2, \\
    \left.\frac{\Delta \gamma ^{E_z}}{\gamma_0}\right\vert_{r_0=0} & = -\frac{1}{4}\varepsilon^2, \ \left.\frac{\Delta \gamma ^{E_z}}{\gamma_0}\right\vert_{r_0=\sigma_0} & = -0.1\varepsilon^2.
    \end{align}
    \label{Eq: Delta_pz_Ez_diff_r0}
\end{subequations}

\subsection{Total energy and momentum shift}
\label{subsec: overall_changes_pz_gamma}

The particle dynamics is determined by the combined action of the Lorentz force $F_\mathrm{L}$ and the induced longitudinal field $E_z$. The resulting total shifts in longitudinal momentum and energy are:
\begin{subequations}
    \begin{align}
     \Delta p_z & = \Delta p_z ^{F_\mathrm{L}} + \Delta p_z ^{E_z}, \label{Eq: overall_Delta_pz} \\
     \Delta \gamma & = \Delta \gamma ^{F_\mathrm{L}} + \Delta \gamma ^{E_z}, \label{Eq: overall_Delta_gamma}
    \end{align}
    \label{Eq: overall_Delta_pz_gamma}
\end{subequations}
where $\Delta p_z ^{F_\mathrm{L}}$ and $\Delta \gamma ^{F_\mathrm{L}}$ are given by Eq. \eqref{Eq: perturbation_solution_Delta}; $\Delta p_z ^{E_z}$ and $\Delta \gamma ^{E_z}$ are given by Eqs. \eqref{Eq: Delta_pz_Ez} and \eqref{Eq: Delta_gamma_Ez}.

Equation \eqref{Eq: overall_Delta_pz} reveals that the momentum loss occurs at all radial positions $r_0$ (with $\Delta p_z < 0$ for the electron beam and $\Delta p_z > 0$ for the positron beam). Notably, while both individual components of the momentum shift ($\Delta p_z ^{F_\mathrm{L}}$ and $\Delta p_z ^{E_z}$) depend strongly on $r_0$, their sum ($\Delta p_z$) is nearly uniform across the beam profile. Peripheral particles ($r_0 = \sigma_0$) experience the maximum momentum loss, which scales as:
\begin{equation}
    \frac{(\Delta p_z)_{\mathrm{max}}}{p_{z,0}} = -0.28\varepsilon^2.
    \label{Eq: max_overall_Delta_pz}
\end{equation}

In contrast, the net energy shift exhibits a clear radial gradient. Due to the energy gain from the Lorentz force at large radii [Eq. \eqref{Eq: Delta_gamma_without_Ez}], peripheral particles experience a net energy gain of $\Delta \gamma /\gamma_0 \simeq 0.1\varepsilon^2$. Conversely, particles near the axis suffer from the lack of Lorentz-driven gain and are dominated by the $E_z$ depletion. Consequently, on-axis particles ($r_0=0$) suffer the maximum energy loss:
\begin{equation}
    \frac{(\Delta \gamma)_{\mathrm{max}}}{\gamma_0} = -\frac{1}{4}\varepsilon^2.
    \label{Eq: max_Delta_gamma_tot}
\end{equation}

\section{Model validation via particle-in-cell simulations}
\label{sec: PIC_simulation_verify_theory}

To validate the theoretical model in Sec. \ref{sec: model}, we perform three-dimensional (3D), fully electromagnetic, particle-in-cell (PIC) simulations using OSIRIS \cite{OSIRIS}. The simulations have employed the beam profile and collision configuration considered by the model (see Sec. \ref{sec: model} and Fig. \ref{fig: schematic_collision_setup_model}). The simulations operate in Cartesian geometry $(x, y, s)$. We fix the beam energy at $\mathcal{E}_0 = 10\ \mathrm{GeV}$ and maintain a constant disruption parameter of $D = 1.88$. By systematically varying the transverse size $\sigma_0$, longitudinal length $\sigma_z$, and density $n_0$, we scan a broad parameter space of $\varepsilon$ spanning over two orders of magnitude ($0.01 \le \varepsilon \le 1.7$). Detailed simulation configurations and the full set of beam parameters are provided in Appendix \ref{SM_sec: PIC_simulation_verify_model}.

\begin{figure}
\includegraphics[width=8.6cm,height=11.2cm]{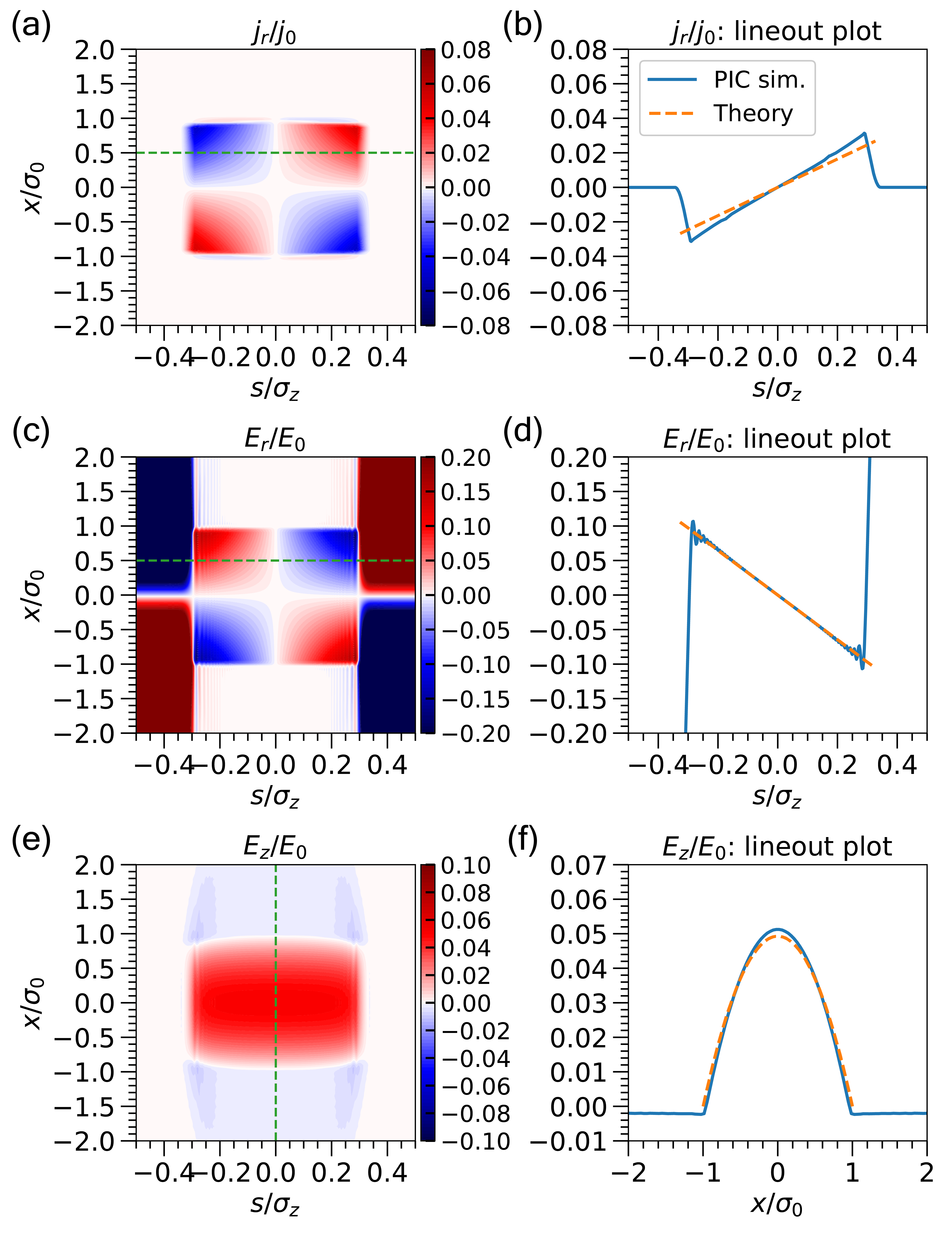}
\caption{(Color online). A snapshot at $t=0.34\tau_{col}$ from a PIC simulation (with OSIRIS) for a collision with $\varepsilon=0.12$. The beam parameters are: $\mathcal{E}_0=10\ \mathrm{GeV}$, $\sigma_0=14.8\ \mathrm{nm}$, $\sigma_z=169.7\ \mathrm{nm}$, and $N_0=4.21\times 10^{9}$. See Sec. \ref{sec: PIC_simulation_verify_theory} for details. Left column: transverse current $j_r/j_0$ in (a) where $j_0 = n_0ec$, transverse field $E_r/E_0$ in (c) where $E_0=2\pi e n_0\sigma_{0}$, and longitudinal field $E_z/E_0$ in (e); Right column: lineout plots along the green dashed lines in the left column.}
\label{fig: current_field_unif_beam}
\end{figure}

A representative case for $\varepsilon = 0.12$ is presented in Fig. \ref{fig: current_field_unif_beam} for a snapshot in the early interaction stage. As illustrated in the lineout plots (right column), the theoretical predictions for the transverse current $\delta j_r$, transverse field $E_r$, and longitudinal field $ E_z$ are in good agreement with the simulation results. $E_z$ remains largely uniform along the $s$-axis, providing a numerical justification for the assumption $\partial \delta E_z / \partial s \simeq 0$ utilized in our analysis in Sec. \ref{subsec: field_Ez}. 

\begin{figure*}
\includegraphics[width=13cm,height=11.2cm]{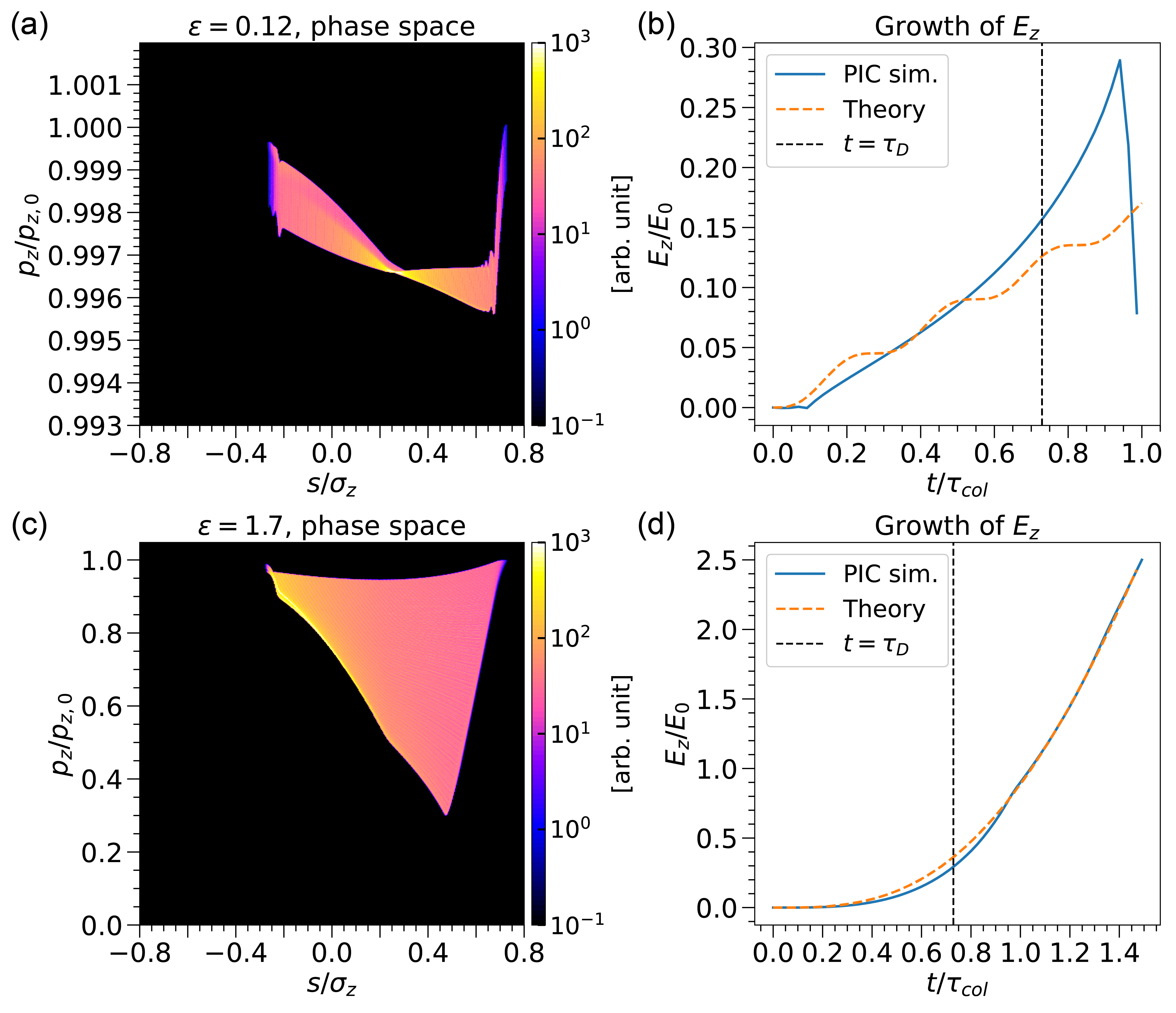}
\caption{(Color online). Left column: phase space ($s, p_z$) of the electron beam (propagating along $+s$) at $t=\tau_D=0.73\tau_{col}$, obtained from PIC simulations for $\varepsilon=0.12$ in (a) (the same case shown in Fig. \ref{fig: current_field_unif_beam}), and $\varepsilon=1.7$ in (c), respectively. Right column: growth of $E_z$ (blue solid), recorded at the interaction center ($r=0$, $s=0$), corresponding to the cases in the left column. The theoretical result [Eq. \eqref{Eq: delta_Ez0}] is shown by the orange dashed line. The vertical line indicates the disruption time scale $\tau_D$.}
\label{fig: phase_Ez_unif_beam}
\end{figure*}

Figure \ref{fig: phase_Ez_unif_beam} shows the phase space of the electron beam at $t=\tau_D$, and the temporal growth of $E_z$. The loss of momentum $p_z$ is observed in Figs. \ref{fig: phase_Ez_unif_beam}(a) and \ref{fig: phase_Ez_unif_beam}(c). The loss of $p_z$ increases rapidly when the collision enters the high-$\varepsilon$ regime ($\varepsilon \ge 1$) [cf. Figs. \ref{fig: phase_Ez_unif_beam}(a) and \ref{fig: phase_Ez_unif_beam}(c)]. The electrons in Fig. \ref{fig: phase_Ez_unif_beam}(c) are still propagating forward at the shown moment ($t=\tau_D$). However, the density pinch and braking effect continue after $t>\tau_D$. A significant fraction of electrons in the case of Fig. \ref{fig: phase_Ez_unif_beam}(c) will lose their entire momentum $p_z$ and then reverse with $p_z<0$, propagating backwards along $-s$ direction. This reversal alters the collision kinematics by prolonging the interaction beyond the nominal collision time ($\tau_{col}=\sigma_z/c$). Consequently, the induced field $E_z$ continues to grow even for $t > \tau_{col}$, as shown in Fig. \ref{fig: phase_Ez_unif_beam}(d). 

\begin{figure}
\includegraphics[width=7.3cm,height=6.7cm]{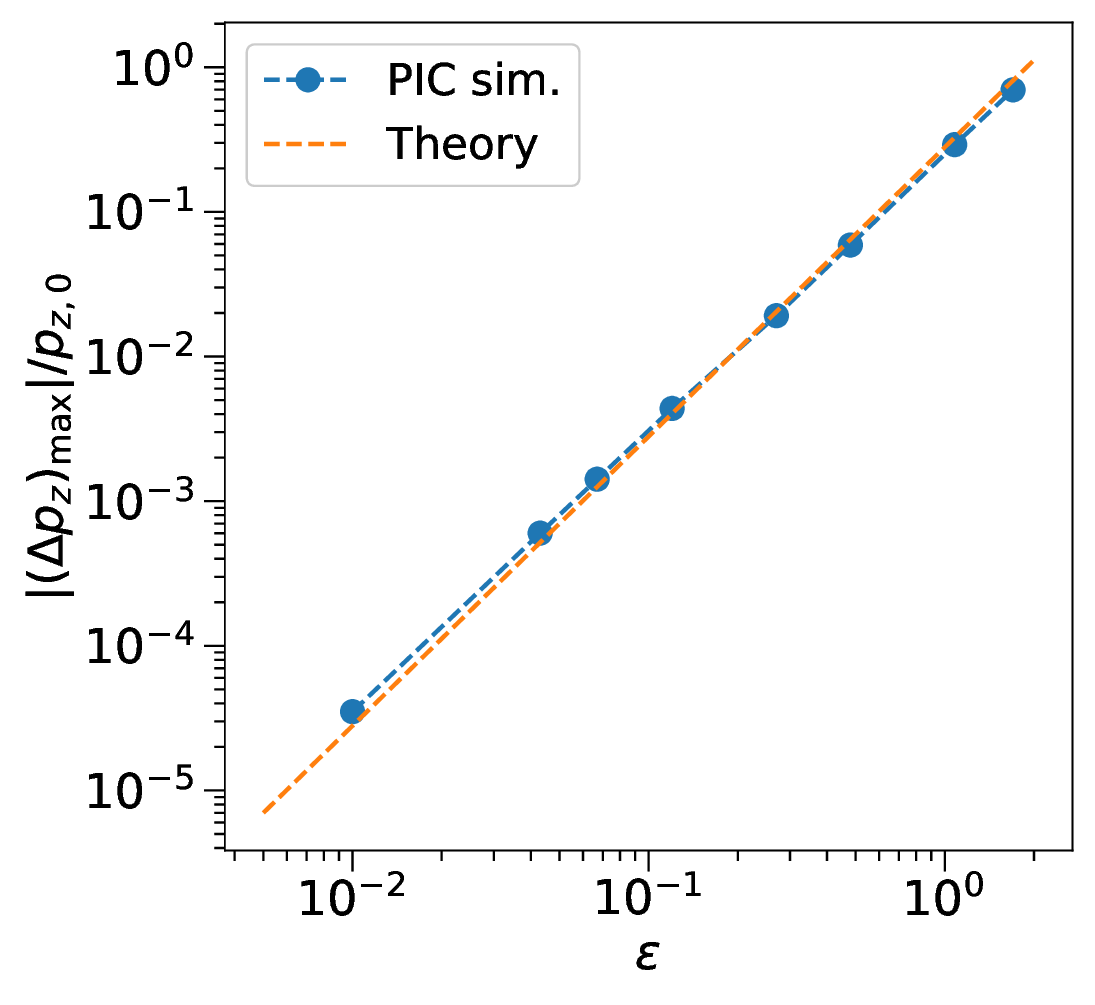}
\caption{(Color online). The maximum longitudinal momentum loss at $t=\tau_D$, i.e., $|(\Delta p_z)_{\mathrm{max}}|/p_{z,0}$, as a function of $\varepsilon$. The symbols show the simulation results (See Sec. \ref{sec: PIC_simulation_verify_theory} and Appendix \ref{SM_sec: PIC_simulation_verify_model} for details), compared with the theoretical prediction [Eq. \eqref{Eq: max_overall_Delta_pz}].}
\label{fig: delta_pz_eps_unif_beam}
\end{figure}

Figure \ref{fig: delta_pz_eps_unif_beam} demonstrates the maximum $p_z$ loss $(\Delta p_z)_{\mathrm{max}}$ as a function of $\varepsilon$. $(\Delta p_z)_{\mathrm{max}}$ is determined by the total time-integrated impulse that combines contribution from both Lorentz force and induced $E_z$. Our model [Eq. \eqref{Eq: max_overall_Delta_pz}] agrees well with the simulations results across the entire range of investigated $\varepsilon$.

In the small-$\varepsilon$ regime, the analytical model for $E_z$ induction [Eq. \eqref{Eq: delta_Ez0}] agrees well with the PIC simulation in the early stage of $t \le 0.5\tau_{col}$, as demonstrated in Fig. \ref{fig: phase_Ez_unif_beam}(b). The oscillatory structure in the theoretical curve in Fig. \ref{fig: phase_Ez_unif_beam}(b) arises from the $\sin$ term in Eq. \eqref{Eq: delta_Ez0}. This structure results from the perturbation method which does not consider the pinch-induced field enhancement and therefore underestimates the particle motion and transverse current $j_r$. In PIC simulations, the particles are driven by realistic fields which are dynamically increasing. This fact smears out the oscillatory structure. The underestimation of $j_r$ also leads to the theoretical $E_z$ which falls below the simulation result when $t$ approaches $\tau_D$ (see Fig. \ref{fig: phase_Ez_unif_beam}(b)). For $t\gtrsim \tau_D$, $E_z$ in the simulation grows even faster, because the beam pinch intensifies which drives more severe current perturbations.

Surprisingly, the analytical model is in a remarkable agreement with PIC results in the high-$\varepsilon$ regime, as shown in Fig. \ref{fig: phase_Ez_unif_beam}(d). This agreement arises from a subtle interplay between our modeling assumptions and the nonlinear beam dynamics. Our derivation assumes that the induced field is insensitive to the longitudinal location ($\partial \delta E_z / \partial s \simeq 0$), an approximation strictly valid for high-aspect-ratio beams ($\sigma_0 \ll \sigma_z$) [cf. Fig. \ref{fig: current_field_unif_beam}(e)]. However, the high-$\varepsilon$ case in Fig. \ref{fig: phase_Ez_unif_beam}(d) is achieved using beams with $\sigma_0 \sim \sigma_z$. In this geometry, $\delta E_z$ develops a significant parabolic-like profile with respect to $s$, where $\partial \delta E_z / \partial s < 0$ for $s > 0$ and $\partial \delta E_z / \partial s > 0$ for $s < 0$. By neglecting this longitudinal gradient in Gauss’s Law [Eq. \eqref{Eq: Gauss_law_satisfaction}], the model intrinsically overestimates $\delta E_r$.

Meanwhile, the severe energy loss in the high-$\varepsilon$ regime enhances the disruption parameter ($D \propto \gamma^{-1}$), leading to stronger density pinch \cite{Zhang2025, Samsonov2021} and field amplification \cite{Zhang2025}. The overestimation of $\delta E_r$ by the model cancels the pinch-induced field amplification. This leads to a good agreement between the model prediction and simulation result of $E_z$ growth for the entire collision course (including $t>\tau_D$), as shown in Fig. \ref{fig: phase_Ez_unif_beam}(d). This  agreement in the high-$\varepsilon$ regime has been verified by our simulations for different beam parameters.

\begin{figure*}
\includegraphics[width=17.26cm,height=6.2cm]{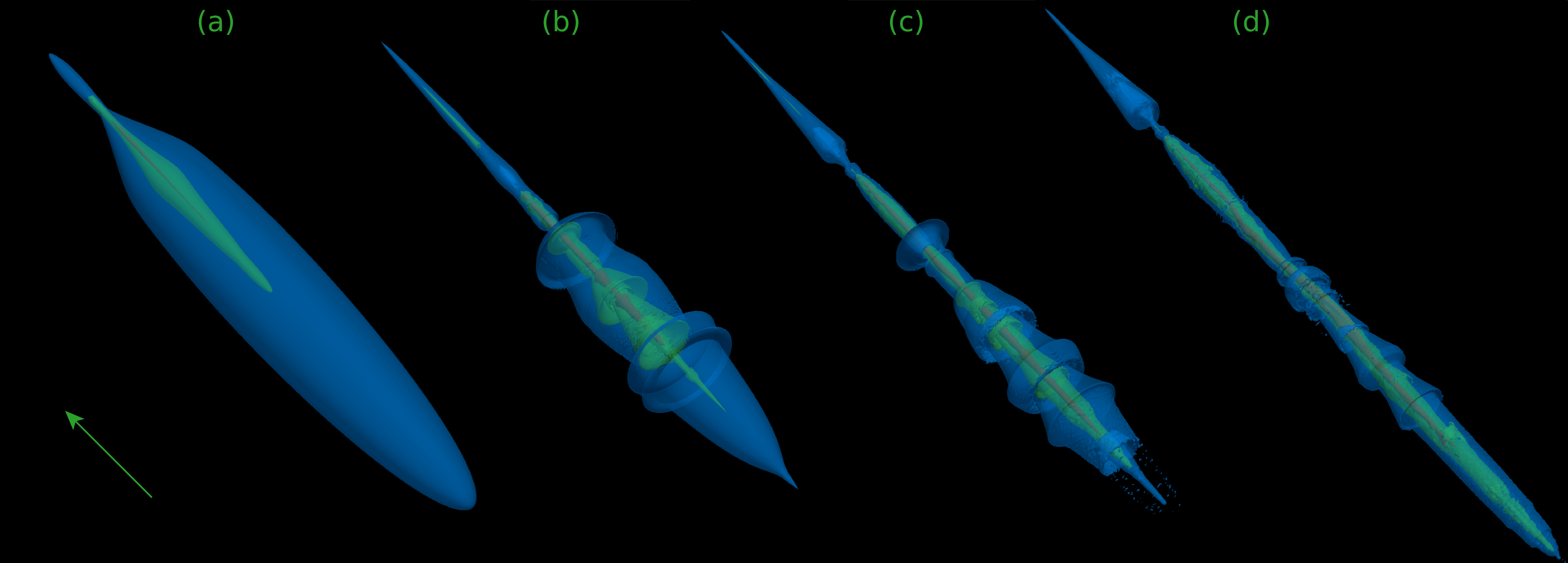}
\caption{(Color online). Visualization of the positron beam from a PIC simulation (with OSIRIS) for a high-$\varepsilon$ collision (with $\varepsilon=3.5$). See Sec. \ref{sec: high_eps_simulations} for details. The simulation does not consider SF-QED processes. Isosurface contours of densities at $n=0.07n_0$ (blue, outermost), $n_0$ (green), and $10n_0$ (red, innermost), respectively, are shown here, at moments of $t=0.36\tau_{col}$ in (a), $0.56\tau_{col}$ in (b), $0.7\tau_{col}$ in (c), and $0.83\tau_{col}$ in (d).}
\label{fig: 3D_beam_fourmoments_eps_2d5}
\end{figure*}

\begin{figure}
\includegraphics[width=8.5cm,height=9.5cm]{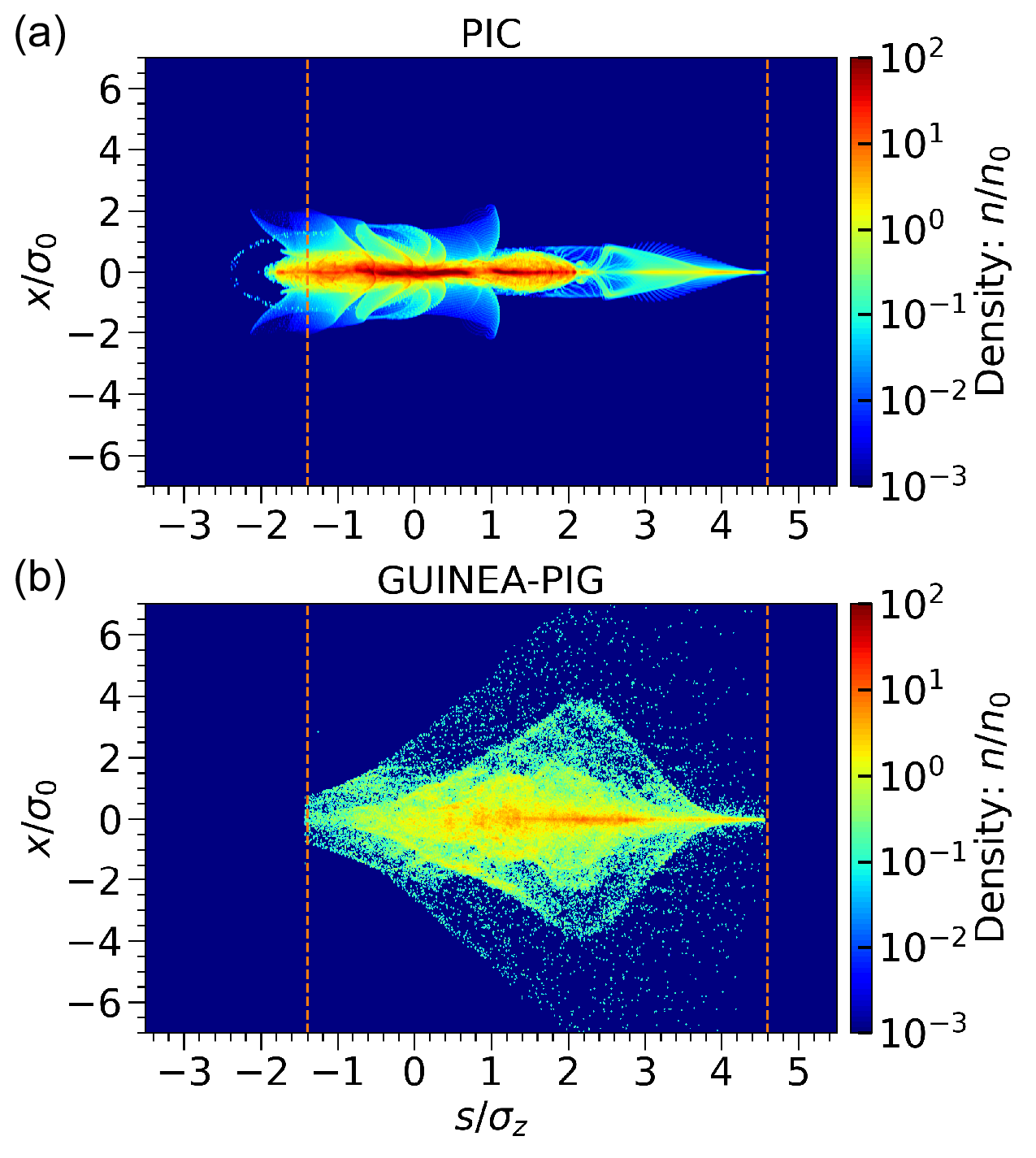}
\caption{(Color online). Density ($n/n_0$) of the (right-moving) electron beam in the high-$\varepsilon$ collision visualized in Fig. \ref{fig: 3D_beam_fourmoments_eps_2d5}. The PIC simulation shown in (a) is compared with GUINEA-PIG simulation illustrated in (b). The density is taken at $t=0.76\tau_{col}$ and extracted from the ($x$, $s$) plane across the beam center. The vertical dashed lines indicate the ``expected'' beam edges assuming that the beam streams at the speed of light.}
\label{fig: 2D_beamslice_eps_2d5}
\end{figure}

\section{PIC and GUINEA-PIG simulations}
\label{sec: high_eps_simulations}

The PIC results are benchmarked against the widely used beam-beam code GUINEA-PIG \cite{Schulte1996}. GUINEA-PIG was originally designed for high-energy colliders operating in the $\varepsilon \ll 1$ regime, where the 2D electrostatic (ES) field solver has been proven highly reliable. By comparing GUINEA-PIG with our fully electromagnetic (EM) PIC simulations, we aim to delineate the physical regimes of applicability of each approach. In particular, this comparison is intended to reveal when the ES approximation must be supplemented by additional physics as $\varepsilon$ increases, and to identify the parameter range beyond which a fully EM treatment becomes necessary for an accurate description of the collision dynamics.

We simulate a collision case with $\varepsilon = 3.5$. The beams are characterized by Gaussian profiles with $\mathcal{E}_0 = 10\ \text{GeV}$, $N_0 = 1.12 \times 10^{12}$ particles per beam, and symmetric dimensions of $\sigma_z = \sigma_0 = 20\ \text{nm}$. $\sigma_0$ and $\sigma_z$ denote the RMS width and length of the Gaussian beams. A finite normalized emittance of $\epsilon_n = 100\ \text{nm}$ is implemented to ensure a realistic beam divergence. The parameters here are chosen to expose the $\varepsilon$-governed physics in the $\varepsilon \ge 1$ regime, not to model any existing or near-term colliders. To isolate the impact of $\varepsilon$-governed dynamics, strong-field quantum electrodynamics (SF-QED) processes, including beamstrahlung and (coherent or incoherent) pair production, are disabled in this comparison. Further details on the simulation configurations are provided in Appendix \ref{SM_sec: PIC_GP_simulation_high_eps}.

The PIC simulation reveals a series of novel beam features, as visualized in Fig. \ref{fig: 3D_beam_fourmoments_eps_2d5}. In this high-disruption regime ($D = 8$), the beams undergo periodic pinching, which forms approximately three distinct ring structures [Fig. \ref{fig: 3D_beam_fourmoments_eps_2d5}(b)]. The number of observed rings scales roughly as $\sim \sqrt{D}$, consistent with the theoretically expected oscillatory behavior of beam envelope during a collision.

As the interaction continues, the energy loss driven by the induced field $E_z$ triggers a non-linear feedback loop. Since the disruption parameter scales as $D \propto \gamma^{-1}$, the rapid $E_z$-caused energy depletion dynamically enhances the focusing strength, causing an intensified beam pinch. This leads to the emergence of multiple ``jellyfish-like" ring structures [Figs. \ref{fig: 3D_beam_fourmoments_eps_2d5}(c) and \ref{fig: 2D_beamslice_eps_2d5}(a)]. The beam dynamics is further complicated by the spatial and temporal variation of the longitudinal deceleration, which creates a highly non-uniform velocity field where shock-like structures can be driven. In this high-$\varepsilon$ regime, a significant fraction of the particles undergo total arrest and subsequent reversal, appearing as a low-density population of scattered particles in the beam tail [Figs. \ref{fig: 3D_beam_fourmoments_eps_2d5}(c) and \ref{fig: 2D_beamslice_eps_2d5}(a)]. Ultimately, this deceleration and momentum redistribution elongate the beams as observed in the final stages of the collision [Figs. \ref{fig: 3D_beam_fourmoments_eps_2d5}(d) and \ref{fig: 2D_beamslice_eps_2d5}(a)].

The GUINEA-PIG simulation [Fig. \ref{fig: 2D_beamslice_eps_2d5}(b)] does not fully capture the longitudinal dynamics described above. Instead of the prolonged pinch and reversal observed in the self-consistent PIC results, the beam in GUINEA-PIG exhibits a rapid transverse expansion immediately following the initial axis-crossing. The anomalously fast transverse motion observed in GUINEA-PIG is an expected consequence of the free-streaming approximation in GUINEA-PIG. This approximation breaks down for high $\varepsilon$.

\begin{figure}
\includegraphics[width=7cm,height=6.5cm]{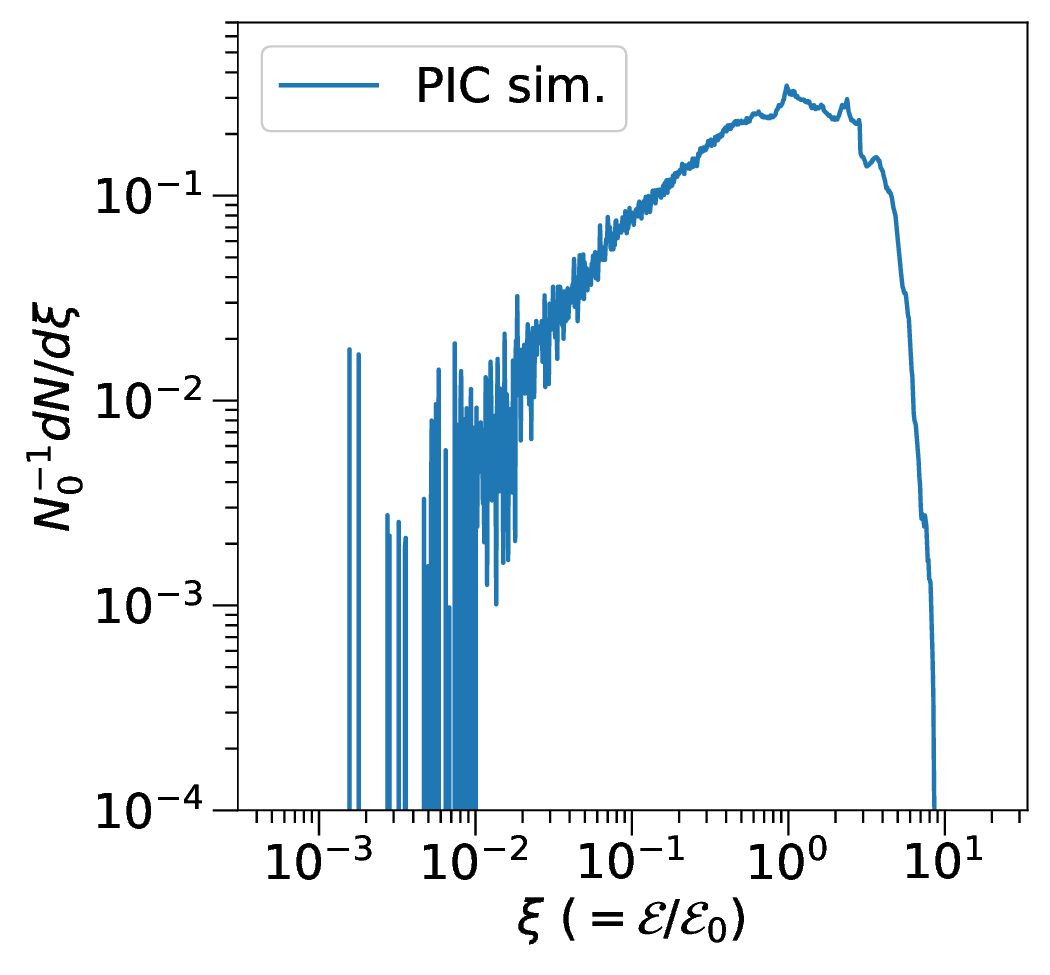}
\caption{(Color online). Energy spectrum $N_0^{-1}dN/d\xi$ for the beam shown in Fig. \ref{fig: 2D_beamslice_eps_2d5}(a) from the PIC simulation, where $\xi = \mathcal{E}/\mathcal{E}_0$.}
\label{fig: ene_spec_eps_2d5}
\end{figure}

The difference in the beam dynamics will impact the collision luminosity. For the case studied here, the PIC-calculated luminosity is $L_0 = 2.2 \times 10^{36}\ \text{cm}^{-2}\text{s}^{-1}$; nearly an order of magnitude higher than the value predicted by GUINEA-PIG. Furthermore, the energy spectrum in Fig. \ref{fig: ene_spec_eps_2d5} clearly exhibits the substantial $E_z$-driven energy loss derived in Secs. \ref{subsec: field_Ez} and \ref{subsec: overall_changes_pz_gamma}.

\begin{figure}
\includegraphics[width=7cm,height=6.5cm]{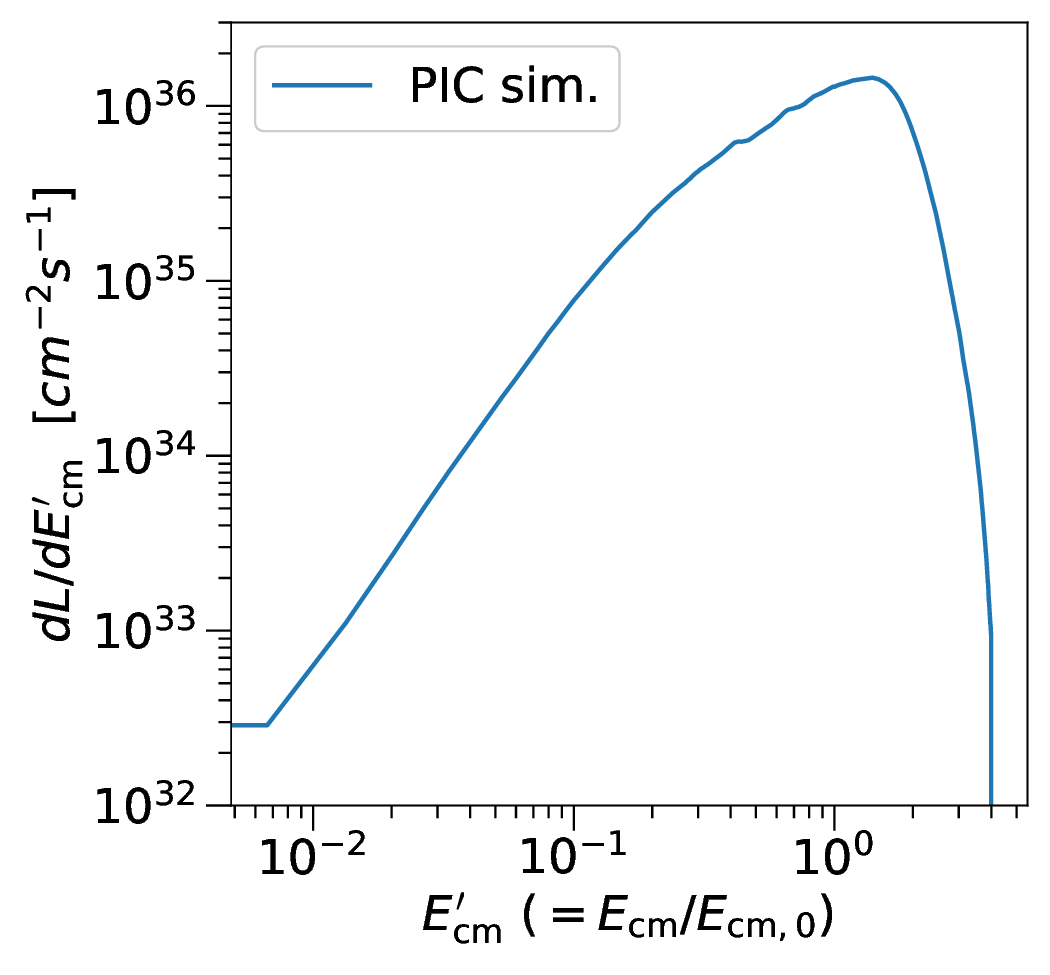}
\caption{(Color online). Luminosity spectrum $dL/dE_{\mathrm{cm}}'$ from the PIC simulation for the high-$\varepsilon$ collision shown in Fig. \ref{fig: 3D_beam_fourmoments_eps_2d5}. $E_{\mathrm{cm}}'=E_{\mathrm{cm}}/E_{\mathrm{cm,0}}$ is the normalized center-of-mass energy, where $E_{\mathrm{cm,0}}=2\mathcal{E}_0$ is the nominal center-of-mass energy.}
\label{fig: spec_lumi_eps_2d5}
\end{figure}

Beyond energy depletion, particles can also gain energy from the radial field $E_r$ during the deflection process [cf. Secs. \ref{subsec: particle_motion_in_collision} and \ref{subsec: analogy_to_electron_motion_EM_wave}], especially for peripheral particles. While the analytical model predicts a net energy gain of $\Delta \gamma / \gamma_0 \simeq 0.1\varepsilon^2$ for the peripheral particles in the $\varepsilon < 1$ regime [Sec. \ref{subsec: overall_changes_pz_gamma}], our PIC simulation reveals that the actual gain can exceed this scaling by a full order of magnitude [see Fig. \ref{fig: ene_spec_eps_2d5}]. This dramatic enhancement is a direct result of the severe beam pinch, which locally amplifies the field $E_r$ \cite{Zhang2025}. A consequence of this process is the emergence of collision events at energies far exceeding the nominal center-of-mass energy ($E_{\text{cm},0} = 2\mathcal{E}_0$). This is illustrated by the luminosity spectrum in Fig. \ref{fig: spec_lumi_eps_2d5}, showing that center-of-mass energies ($E_{\text{cm}}$) can reach approximately $4E_{\text{cm},0}$. The technical definitions of the luminosity spectrum and the associated data processing methodology are detailed in Appendix \ref{SM_sec: luminosity_spectrum}.

\begin{figure}
\includegraphics[width=7cm,height=6.6cm]{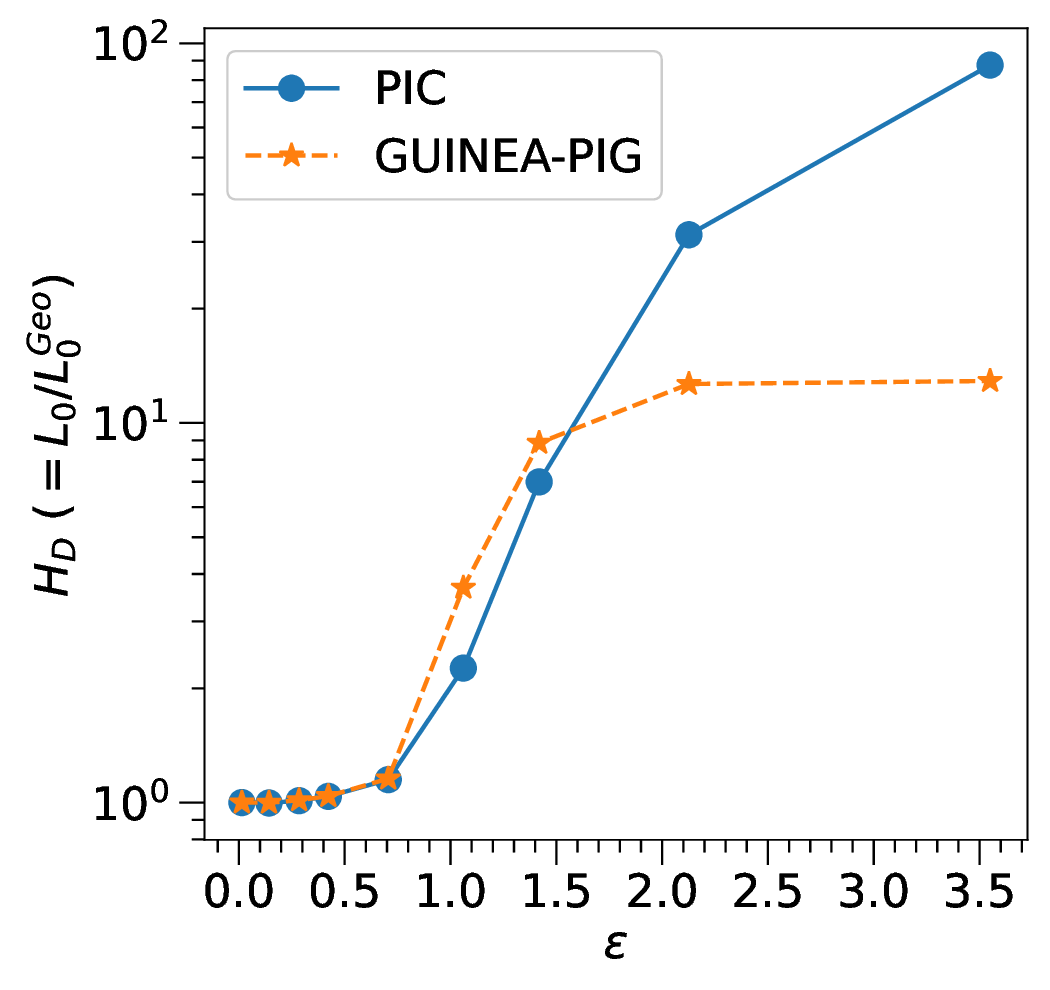}
\caption{(Color online). Luminosity enhancement $H_D$ as a function of $\varepsilon$, where $H_D = L_0/L_0^{Geo}$, with $L_0^{Geo}=N_0^2/(4\pi \sigma_0^2)$ the geometric luminosity for a Gaussian-beam collision \cite{Zhang2025}. The symbols represent the PIC and GUINEA-PIG simulations, respectively. The particle number $N_0$ is varied to render different $\varepsilon$, and other beam parameters remain the same as those of the collision shown in Fig. \ref{fig: 3D_beam_fourmoments_eps_2d5} (Sec. \ref{sec: high_eps_simulations}). SF-QED processes are disabled here.}
\label{fig: HD_eps_NoQED}
\end{figure}

The scaling of total luminosity with the parameter $\varepsilon$ is investigated in Fig. \ref{fig: HD_eps_NoQED}, revealing a fundamental divergence between the two simulation frameworks. While the PIC-calculated luminosity grows rapidly with $\varepsilon$ as the interaction enters the high-$\varepsilon$ regime, GUINEA-PIG exhibits a saturation. The PIC simulations give higher luminosities, because the beam pinch is enhanced, since $\gamma$ is decreased by the field $E_z$. The beams are eventially elongated due to particle reversal demonstrated in this section. For very large $\varepsilon$, we expect $H_D$ in PIC simulations to saturate and plateau with increasing $\varepsilon$ since the particle reversal occurs so fast that the effective collision rate begins to decrease.

\section{Discussion}
\label{sec: discussion}

In this section, we explore the physical significance of $\varepsilon$. We first show that it determines whether the beam’s self-fields dominate the collision dynamics, as it directly measures the ratio of self-field energy density to beam kinetic energy density. We then demonstrate that analogous parameters arise in other physical scenarios involving axial field induction, taking finite-spot laser focusing in vacuum as an illustrative example. Another key issue is the interplay between the strong-field QED (SF-QED) processes and the dynamics governed by $\varepsilon$ proposed here. SF-QED processes in beam collisions include beamstrahlung (photon emission) and electron-positron pair production \cite{yokoya2005beam, Fabrizio2019, Zhang2023, Zhang2025}. The coherent pair production dominates at large $\chi$. The PIC simulations with OSIRIS shown in this study have included the coherent pair production. Nevertheless, these two mechanisms act as competing channels through which particles lose energy and momentum, and their interplay gives rise to a nonlinear coupling that significantly modifies the collective beam dynamics and the collision luminosity. We finally consider the relevant indications of our study for the future high-energy $e^-e^+$ linear colliders. The intense beam-beam effects in plasma-based colliders can cause very-low-energy particles which may undergo large-angle deflections or complete back-scattering.

\emph{Energy perspective to $\varepsilon$}---The kinetic energy density of a beam is
\begin{equation}
    u_0=n_0\gamma_0mc^2.
    \label{Eq: energy_density_u0}
\end{equation}
The energy density of the self-fields is 
\begin{equation} 
    u_{\mathrm{f}}=\frac{1}{8\pi}(E_r^2+B_\theta^2) \simeq \frac{1}{4\pi}E_0^2.
\end{equation}
A surprising relation can be found by comparing these two energy densities, i.e.,
\begin{equation}
    \frac{u_{\mathrm{f}}}{u_0} \simeq \frac{1}{4}\varepsilon^2.
    \label{Eq: uf_u0_eps}
\end{equation}
Equation \eqref{Eq: uf_u0_eps} shows that the ratio of self-fields to kinetic energy density scales as $\varepsilon^2$. Consequently, a large-angle disruption ($\varepsilon \ge 1$) amounts to a field-dominated beam-beam interaction.

\emph{``Equivalent $\varepsilon$-like parameter" for a laser focusing in vacuum}---When a laser with a finite spot size is focused in vacuum, it generates an axial electric field $E_{z,\mathrm{L}}$ \cite{Esarey2009}
\begin{equation}
    E_{z,\mathrm{L}}\sim \frac{r_{0,\mathrm{L}}}{Z_R}E_{\perp,\mathrm{L}},
    \label{Eq: Ez_laser}
\end{equation}
where $r_{0,\mathrm{L}}$ is the spot size (beam waist) at focus, $Z_R$ is the Rayleigh length, and $E_{\perp,\mathrm{L}}$ is the laser field. 

In a beam-beam collision, each beam acts like a lens and an axial field $E_z$ is also generated, scaling as $E_z \sim \varepsilon E_0$ [Eq. \eqref{Eq: delta_Ez0}]. We can draw an analogy between the beam disruption and the laser propagation in vacuum in terms of axial field growth. The coefficient $\frac{r_{0,\mathrm{L}}}{Z_R}$ in Eq. \eqref{Eq: Ez_laser} can be considered an ``equivalent $\varepsilon$-like parameter," since it has similar definition and role compared to $\varepsilon$ (cf. Eq. \eqref{Eq: eps_definition}). Both parameters represent the ratio between the transverse size and the characteristic longitudinal dynamical length.

\emph{Impact of SF-QED}---We begin by demonstrating how SF-QED processes impact the $\varepsilon$-governed dynamics. For this goal, we need to isolate SF-QED from other relevant factors. To do so, we conduct a controlled case study in the regime of initially small $\varepsilon$ ($\varepsilon < 1$) and negligible disruption ($D \ll 1$). By choosing these parameters, we ensure that any significant changes in the beam evolution can be directly attributed to the energy loss from SF-QED. A brief discussion for the distinction between SF-QED and the $\varepsilon$-governed dynamics is provided in Appendix \ref{SM_sec: distinction_SFQED_epsdynamics}. Furthermore, we maintain a regime where pair production remains a minor perturbation relative to the primary beam densities, thereby avoiding the complex back-reactions on the self-fields that characterize the massive-pair-production regime \cite{Zhang2025}.

For consistency, we maintain the beam geometry and energy used in the high-$\varepsilon$ analysis in Sec. \ref{sec: high_eps_simulations} ($\mathcal{E}_0 = 10\ \text{GeV}$, $\sigma_0 = \sigma_z = 20\ \text{nm}$), but reduce the particle number to $N_0 = 7.2 \times 10^9$, resulting in an initial state characterized by $\varepsilon = 0.28$ and $D = 0.05$. The system resides in a strong-quantum regime with $\chi = 276$, where significant SF-QED are expected; according to the scalings in Ref. \cite{Zhang2023}, each particle will emit an average of $n_\gamma \simeq 1.5$ photons. Our PIC simulations with SF-QED have included beamstrahlung and coherent pair production. This setup allows us to observe how beamstrahlung energy depletion acts as a precursor to the large-angle dynamics. Further simulation details are provided in Appendix \ref{SM_sec: PIC_GP_SFQED_simulation}.

\begin{figure}
\includegraphics[width=8.5cm,height=7.8cm]{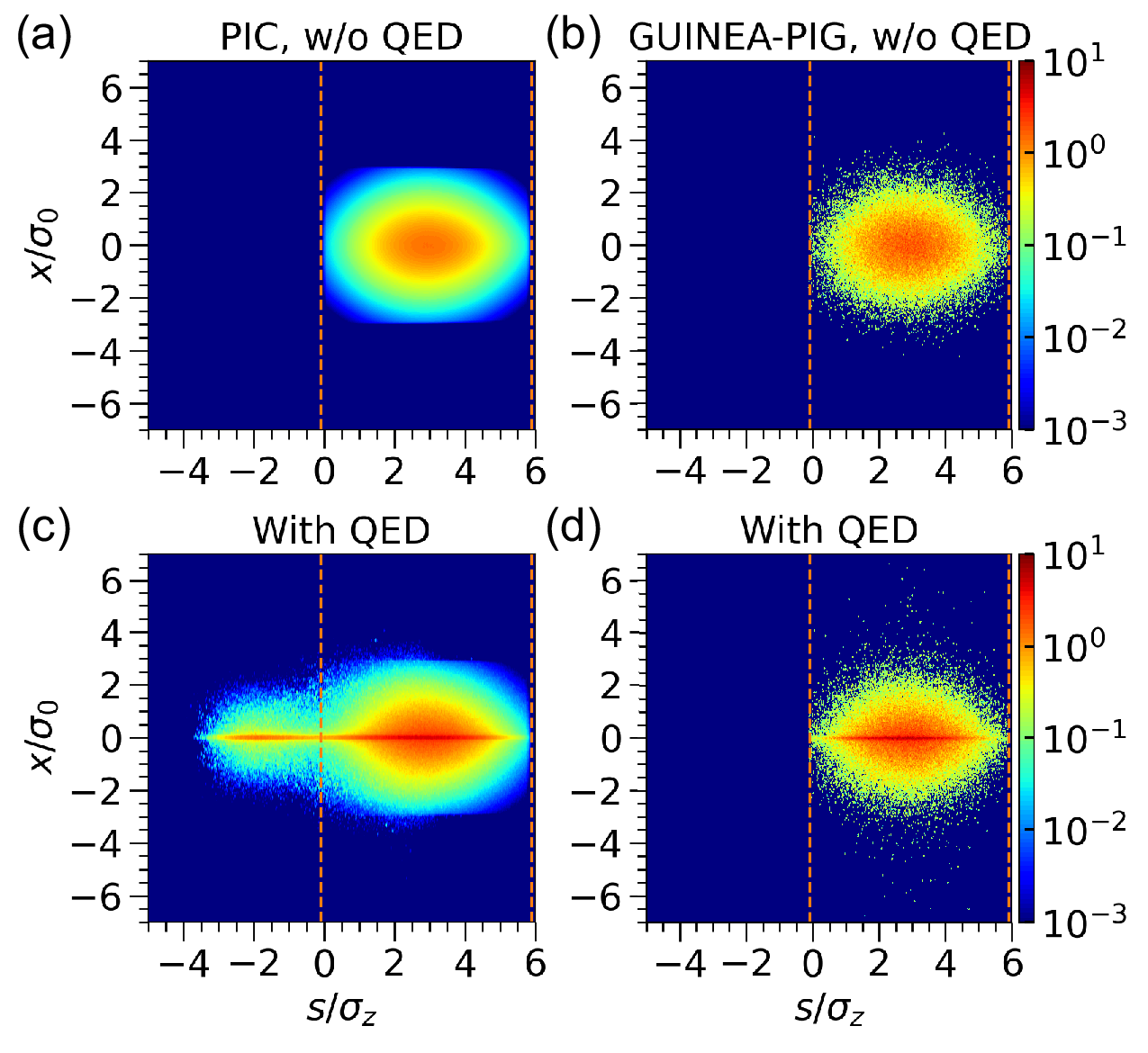}
\caption{(Color online). Density ($n/n_0$ for the color scale) of the (right-moving) electron beam in a low-$\varepsilon$ collision ($\varepsilon=0.28$), at the nominal final moment ($t=\tau_{col}$). Further details can be found in Sec. \ref{sec: discussion} and Appendix \ref{SM_sec: PIC_GP_SFQED_simulation}. (a) and (c) show the PIC simulations w/o and with SF-QED processes, respectively. (b) and (d) show the counterpart GUINEA-PIG simulations. The two vertical dashed lines indicate the ``expected'' beam edges assuming that the beam streams at the speed of light.}
\label{fig: 2D_beams_SFQED_eps_0d2}
\end{figure}

\begin{figure}
\includegraphics[width=7.7cm,height=12cm]{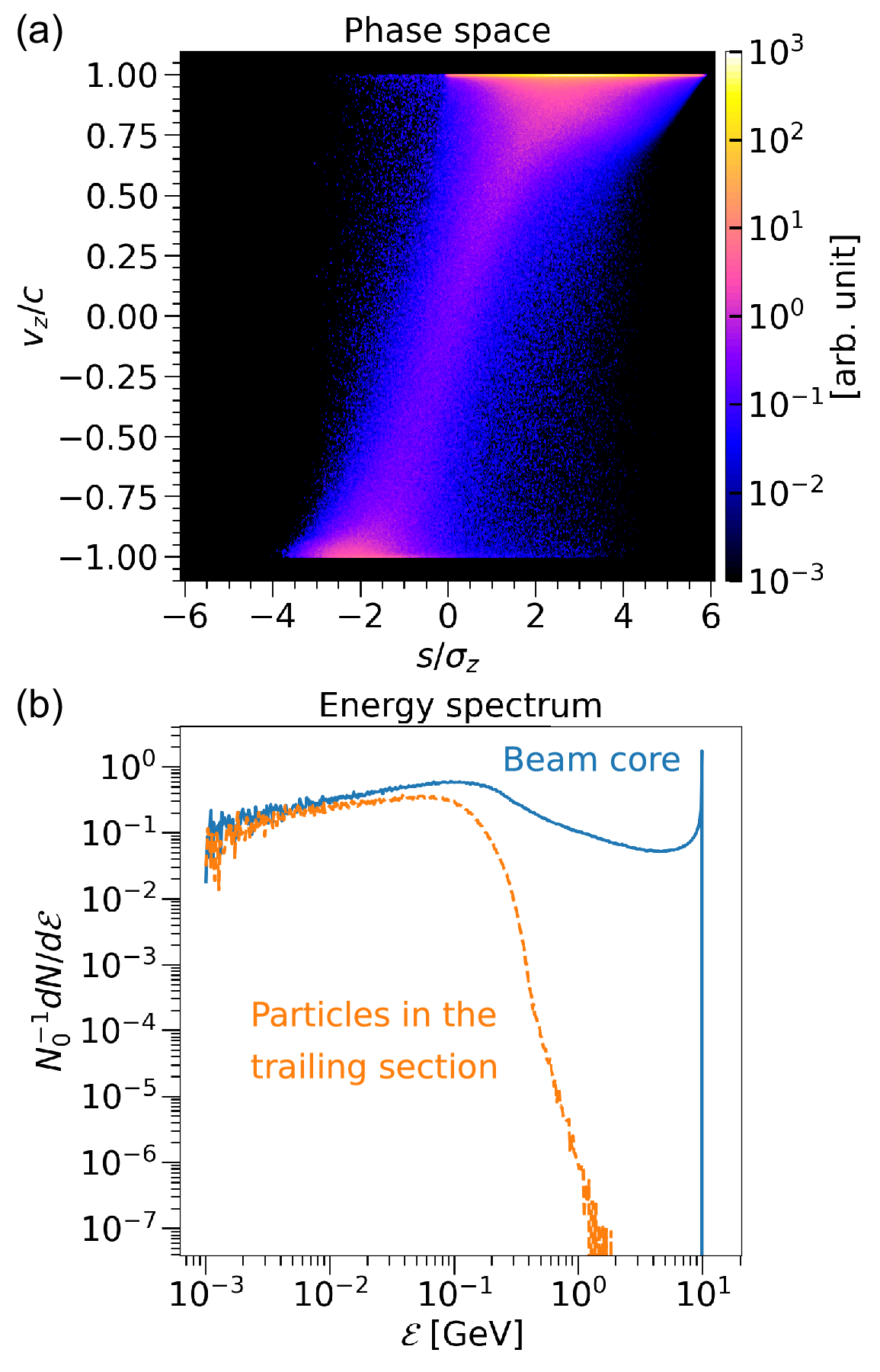}
\caption{(Color online). Analysis for the beam shown in Fig. \ref{fig: 2D_beams_SFQED_eps_0d2}(c) from PIC simulation with SF-QED processes. (a) Phase space ($s, v_z$). (b) Energy spectrum $N_0^{-1}dN/d\mathcal{E}$. ``Beam core" represents the bulk of beam which remains within the two vertical dashed lines in Fig. \ref{fig: 2D_beams_SFQED_eps_0d2}(c); ``Particles in the trailing section" refer to those with coordinates $s<0$, i.e., left behind the ``beam core".}
\label{fig: vz_s_ene_spec_SFQED_eps_0d2}
\end{figure}

Figure \ref{fig: 2D_beams_SFQED_eps_0d2} depicts the electron beam at the nominal end of the collision ($t=\tau_{col}$). In the absence of SF-QED, the beams exhibit minimal perturbations, as both $D$ and $\varepsilon$ are low. The results from the PIC and GUINEA-PIG simulations are quite similar, as seen in the upper row. When SF-QED is activated, beamstrahlung causes energy loss. This intensifies the transverse deflection and raises both $D$ and $\varepsilon$. The transverse current $j_r$ increases significantly (by an order of magnitude here). Moreover, the increase in $\varepsilon$ promotes a stronger braking effect and $E_z$. This impact of SF-QED is corroborated by the PIC simulation. 

A significant fraction of the beam is left behind due to the sustained and enhanced deceleration, forming a trailing section visible to the left of the first vertical dashed line in Fig.~\ref{fig: 2D_beams_SFQED_eps_0d2}(c). As revealed by the corresponding phase space in Fig.~\ref{fig: vz_s_ene_spec_SFQED_eps_0d2}(a), approximately $81\%$ of particles in this trailing section, comprising both primary electrons and newly produced pairs, are moving backward ($v_z < 0$). These reversed particles have suffered severe energy loss, with $\mathcal{E} < 0.4\ \mathrm{GeV}$, as shown by the energy spectrum in Fig.~\ref{fig: vz_s_ene_spec_SFQED_eps_0d2}(b). The energy and momentum losses are initially driven by beamstrahlung, but as $\varepsilon$ increases, the $E_z$-induced losses become dominant, with the strongest deceleration occurring near the axis where $E_z$ is maximized. The late-stage dynamics ($t \gtrsim \tau_{col}$) lie beyond the scope of this study.

\begin{figure}
\includegraphics[width=7cm,height=6.4cm]{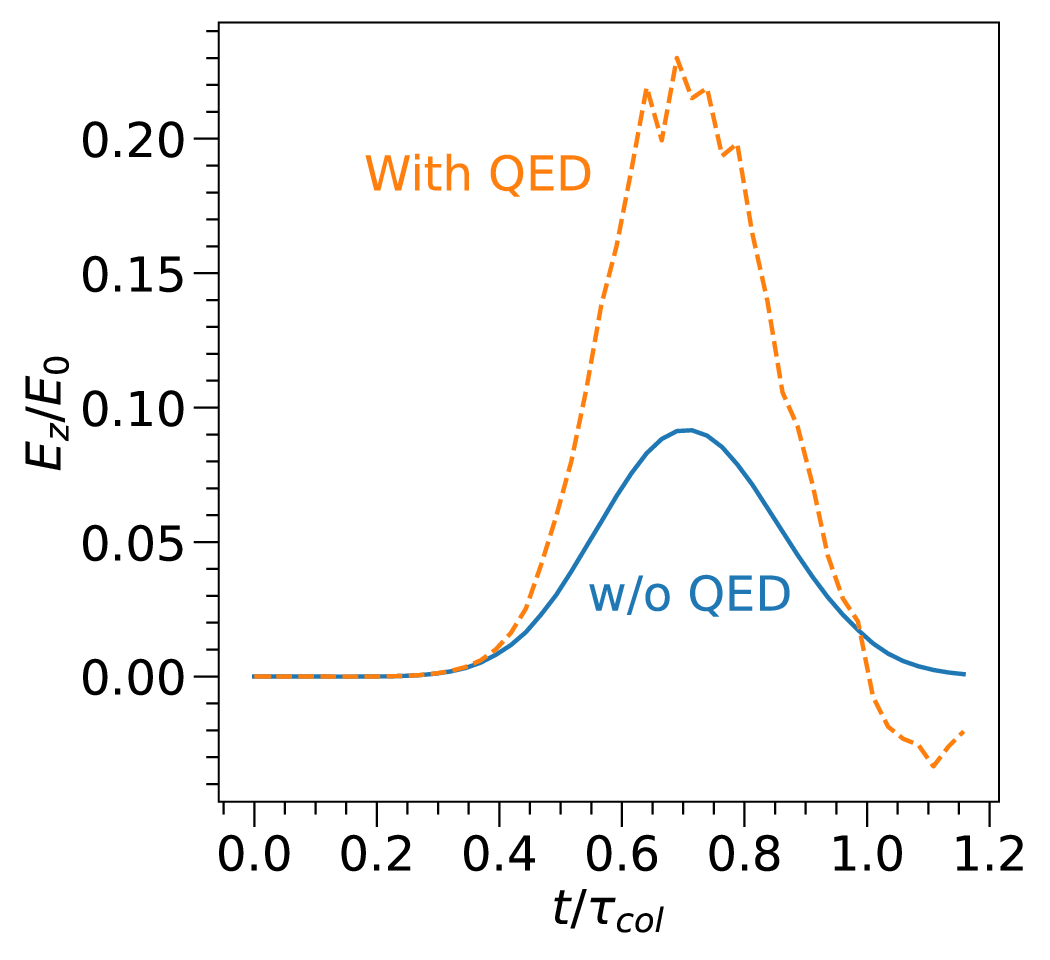}
\caption{(Color online). Growth of $E_z$ at the collision center ($r=0$, $s=0$), recorded in the PIC simulations for the left column of Fig. \ref{fig: 2D_beams_SFQED_eps_0d2}. Here, $E_0$ is the maximum electric field of the Gaussian-profile beams used in the collision.}
\label{fig: Ez_SFQED_eps_0d2}
\end{figure}

GUINEA-PIG assumes free-streaming particles, and thus it cannot capture this deceleration [Fig. \ref{fig: 2D_beams_SFQED_eps_0d2}(d)]. The enhancement of $E_z$ resulting from SF-QED is presented in Fig. \ref{fig: Ez_SFQED_eps_0d2}, showing that $E_z$ with SF-QED is approximately three times stronger than without SF-QED. Additionally, the amplitude of $E_z$ without SF-QED aligns well with our theoretical prediction. 

The process of pair production can play an important role in increasing $\varepsilon$. On one hand, newly generated pairs possess much lower energies, making them more susceptible to deflection \cite{Zhang2023, Zhang2025}. On the other hand, these new electrons and positrons travel in opposite directions within the fields, resulting in a stronger transverse current \cite{Zhang2025}. However, the collision scenario involving substantial pair production falls outside the scope of this study.

\begin{figure*}
\includegraphics[width=12cm,height=8.3cm]{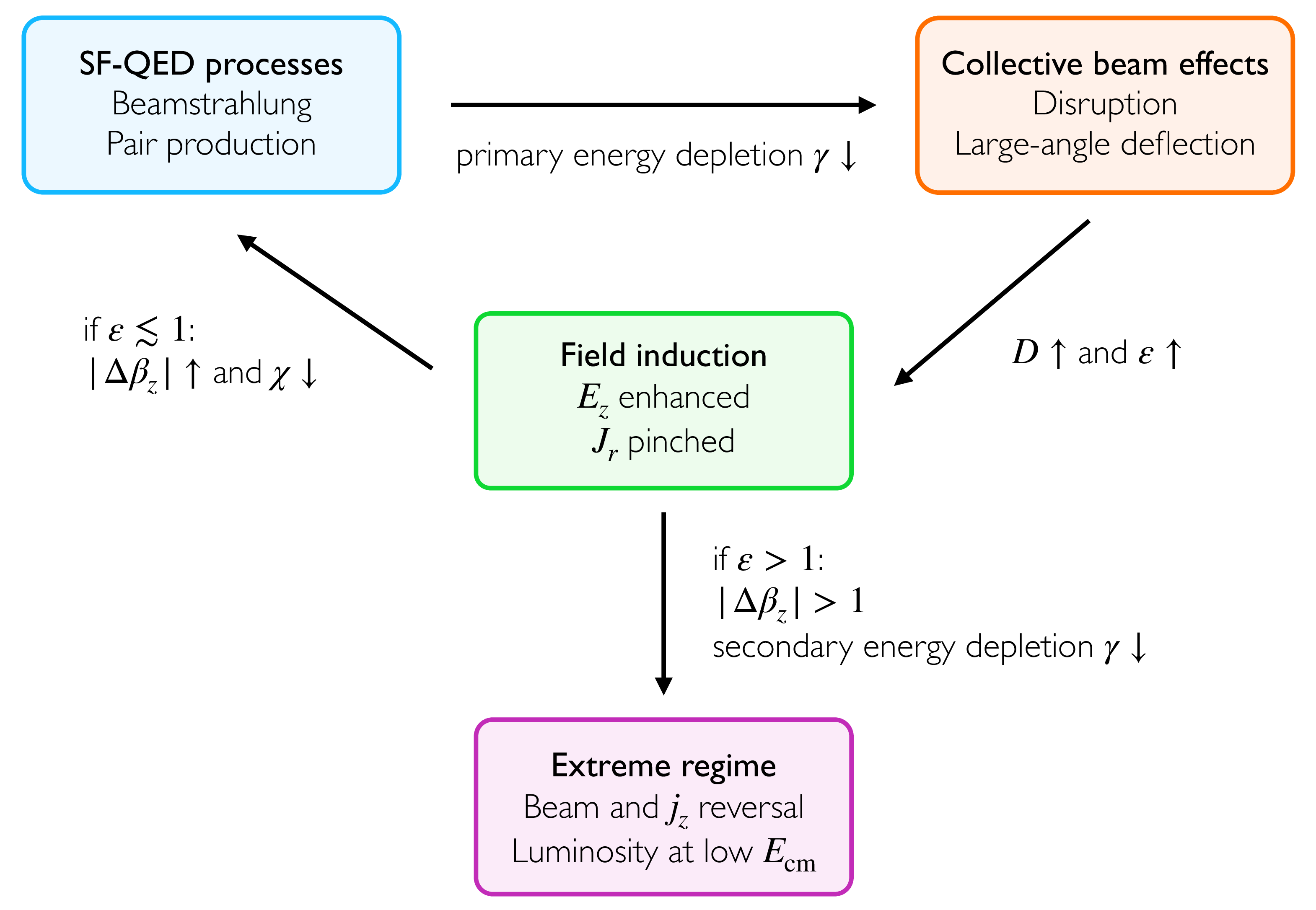}
\caption{(Color online). Schematic diagram of the interplay between SF-QED processes and $\varepsilon$-governed dynamics proposed in this study. A detailed discussion can be found in Sec. \ref{sec: discussion}.}
\label{fig: schematic_SFQED_vs_eps}
\end{figure*}

\emph{Interplay between SF-QED and $\varepsilon$-governed dynamics}---These two mechanisms can couple with each other, featuring an interplay which changes the collective beam dynamics. We introduce another dimensionless parameter, $\kappa$, to characterize this interplay. This parameter $\kappa$ quantifies the comparative rate of energy depletion through these two channels in the early phase of a collision ($t\ll \tau_D$), i.e.,
\begin{equation}
    \kappa = \frac{\mathcal{P}_\varepsilon}{\mathcal{P}_{\mathrm{rad}}},
    \label{Eq: kappa_definition}
\end{equation}
where $\mathcal{P}_\varepsilon$ represents the equivalent power for energy loss through the $\varepsilon$-governed dynamics, and $\mathcal{P}_{\mathrm{rad}}$ denotes the beamstrahlung power. Since the energy loss via the $\varepsilon$-governed dynamics scales as $\Delta \gamma / \gamma_0 \sim -\varepsilon^2$ [Eq. \ref{Eq: max_Delta_gamma_tot}], $\mathcal{P}_\varepsilon$ can be approximately evaluated as
\begin{equation}
    \mathcal{P}_\varepsilon \sim \frac{\varepsilon^2 \mathcal{E}_0}{\tau_D} = \frac{m\omega_b^3\sigma_0^2}{\sqrt{\gamma_0}}.
    \label{Eq: Delta_gamma_power}
\end{equation}
For $\mathcal{P}_{\mathrm{rad}}$, we employ the asymptotic radiation power of a single particle \cite{Gonoskov2022}, i.e.,
\begin{subequations}
\begin{align}
\mathcal{P}_{\mathrm{rad}} & = 0.37\frac{\alpha mc^3}{\lambdabar_\mathrm{c}}\chi^{2/3} \ \ \mathrm{for}\ \chi \gg 1, \label{Eq: subeq_P_rad_large_chi} \\
& = \frac{2}{3} \frac{\alpha mc^3}{\lambdabar_\mathrm{c}}\chi^{2} \ \ \ \ \ \ \  \mathrm{for} \ \ \chi \ll 1, \label{Eq: subeq_P_rad_small_chi}
\end{align}
\label{Eq: P_rad}
\end{subequations}
where $\lambdabar_\mathrm{c}=\hbar/mc$ is the reduced Compton length and $\hbar$ is the reduced Planck constant; $\alpha=e^2/\hbar c$ is the fine-structure constant. With Eqs. \eqref{Eq: Delta_gamma_power} and \eqref{Eq: P_rad}, one has
\begin{subequations}
\begin{align}
\kappa & = 19.5 \left(\frac{\lambdabar_\mathrm{c}}{\gamma_0 \sigma_0}\right)^{1/3} \left(\frac{N_0\lambdabar_\mathrm{c}}{\gamma_0 \sigma_z}\right)^{5/6} \ \ \ \mathrm{for}\ \chi \gg 1, \label{Eq: subeq_kappa_large_chi} \\
& = 0.75 \frac{\sigma_0}{\gamma_0^2 r_e} \left(\frac{\sigma_z}{\gamma_0N_0r_e}\right)^{1/2} \ \ \ \ \ \ \ \ \  \mathrm{for}\ \chi \ll 1, \label{Eq: subeq_kappa_small_chi}
\end{align}
\label{Eq: kappa_results}
\end{subequations}
where $r_e=e^2/mc^2$ is the classical electron radius. In the strong-quantum regime ($\chi \gg 1$), $\kappa$ has an almost linear scaling with the particle number per unit length, i.e., $\kappa \propto (N_0/\sigma_z)^{5/6}$. This suggests that collisions with a high ratio of $N_0/\sigma_z$ will favor the dynamics governed by $\varepsilon$, and this is consistent with the formula of $\varepsilon$ [Eq. \eqref{Eq: eps_engineering_definition}]. However, in any realistic collider design, whether utilizing cigar-shaped or spherical beams, the parameter $\kappa$ typically remains small, generally with $\kappa < 0.01$. It is important to note that if the parameters $D$ and $\varepsilon$ can be dynamically adjusted during the interaction (e.g., due to beamstrahlung), $\kappa$ also varies accordingly. Our PIC simulations support this reasoning, indicating that $\varepsilon$-governed dynamics start to become significant relative to beamstrahlung for $\kappa > 0.01$. A full parameter survey for $\kappa$ will be reported in future work.

\begin{figure}
\includegraphics[width=7cm,height=15cm]{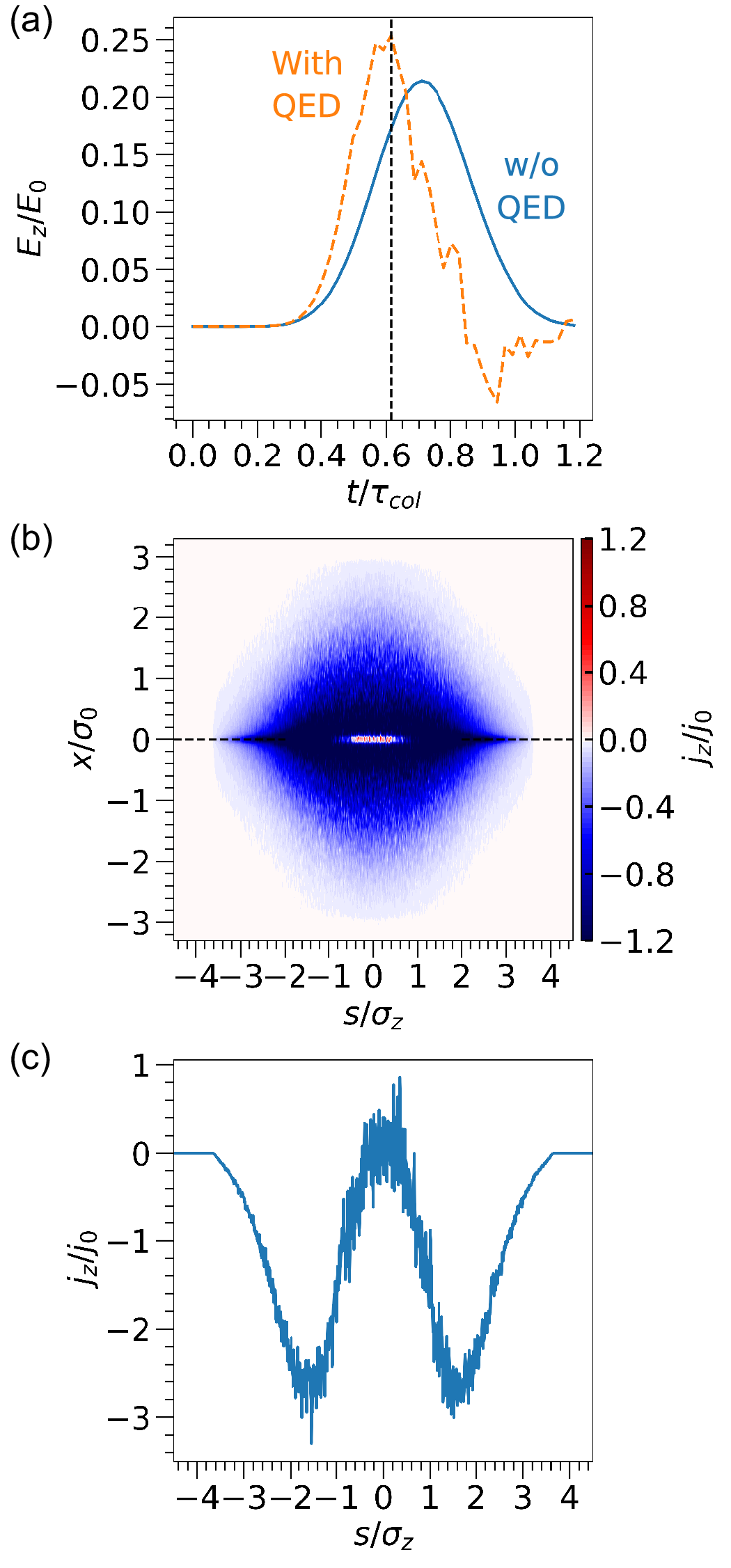}
\caption{(Color online). PIC simulation for a collision which features an interplay between SF-QED and $\varepsilon$-governed dynamics. The beams have the same parameters as those in Fig. \ref{fig: 2D_beams_SFQED_eps_0d2}, but with a higher particle number of $N_0=1.6\times 10^{10}$, leading to moderate $\varepsilon$ ($\varepsilon = 0.42$) and prominent coupling parameter of $\kappa=0.19$. (a) $E_z$ growth. (b) Longitudinal current map ($j_z/j_0$) in the $x$-$s$ plane, at the moment $t=0.61\tau_{col}$ which is indicated by the vertical dashed line in (a). (c) Lineout plot of $j_z/j_0$ shown in (b), along the propagation axis ($x=0$) indicated by the horizontal dashed line in (b). }
\label{fig: Ez_j1_SFQED_eps_0d3}
\end{figure}

For even higher $\kappa$ ($\kappa \gtrsim 0.1$), the $\varepsilon$-governed dynamics become prominent and couple with SF-QED. Their interaction is summarized in Fig. \ref{fig: schematic_SFQED_vs_eps}. On one hand, SF-QED enhances the collective beam effects, including disruption and $\varepsilon$-governed dynamics, as demonstrated in Figs. \ref{fig: 2D_beams_SFQED_eps_0d2}(a) and \ref{fig: 2D_beams_SFQED_eps_0d2}(c). With increased $\varepsilon$, the beam pinch leads to higher current $j_r$ and stronger field $E_z$ [Fig. \ref{fig: Ez_SFQED_eps_0d2}]. On the other hand, the enhanced $\varepsilon$-governed dynamics will, in turn, limit SF-QED. The amplified braking effect leads to a reduction in both the velocity $v_z$ and the perpendicular Lorentz force $F_{\mathrm{L, \perp}}$. Consequently, the quantum parameter $\chi$ decreases, which weakens the effects of SF-QED. If $\varepsilon$ becomes sufficiently large ($\varepsilon \ge 1$), the excessive deceleration can reverse the beam propagation ($|\Delta \beta_z|>1$). The reversed particles will no longer experience beamstrahlung since they co-propagate with the opposing beam. This condition will further invert the orientation of current $j_z$ (demonstrated in Fig. \ref{fig: Ez_j1_SFQED_eps_0d3}). 

We have verified with PIC simulations the interplay between SF-QED and the $\varepsilon$-governed dynamics, as illustrated in Fig. \ref{fig: Ez_j1_SFQED_eps_0d3} for a scenario where $\varepsilon = 0.42$ and $\kappa = 0.19$. The beam pinch leads to strong currents $j_z$ (dark blue parts in Fig. \ref{fig: Ez_j1_SFQED_eps_0d3}(b)) around the propagation axis (horizontal dashed line). The amplitudes of those pinched currents ($\sim -3j_0$) have exceeded the range of color scale ($\pm 1.2j_0$) set here. They are more clearly shown by the lineout plot in Fig. \ref{fig: Ez_j1_SFQED_eps_0d3}(c). A positive current emerges (red region around the center in Fig. \ref{fig: Ez_j1_SFQED_eps_0d3}(b) or the positive part in Fig. \ref{fig: Ez_j1_SFQED_eps_0d3}(c)), as the local beam's propagation is reversed. As a result of the flipped $j_z$, the field $E_z$ near the center begins to decrease and eventually turns negative, as shown in Fig. \ref{fig: Ez_j1_SFQED_eps_0d3}(a) (refer to the moments after the vertical dashed line). These results indicate that SF-QED will not infinitely amplify $\varepsilon$ and $E_z$, since the collective dynamics of beams and fields will ultimately change, leading to a limitation of SF-QED due to the feedback loop identified in this study.

\begin{figure}
\includegraphics[width=7cm,height=12.6cm]{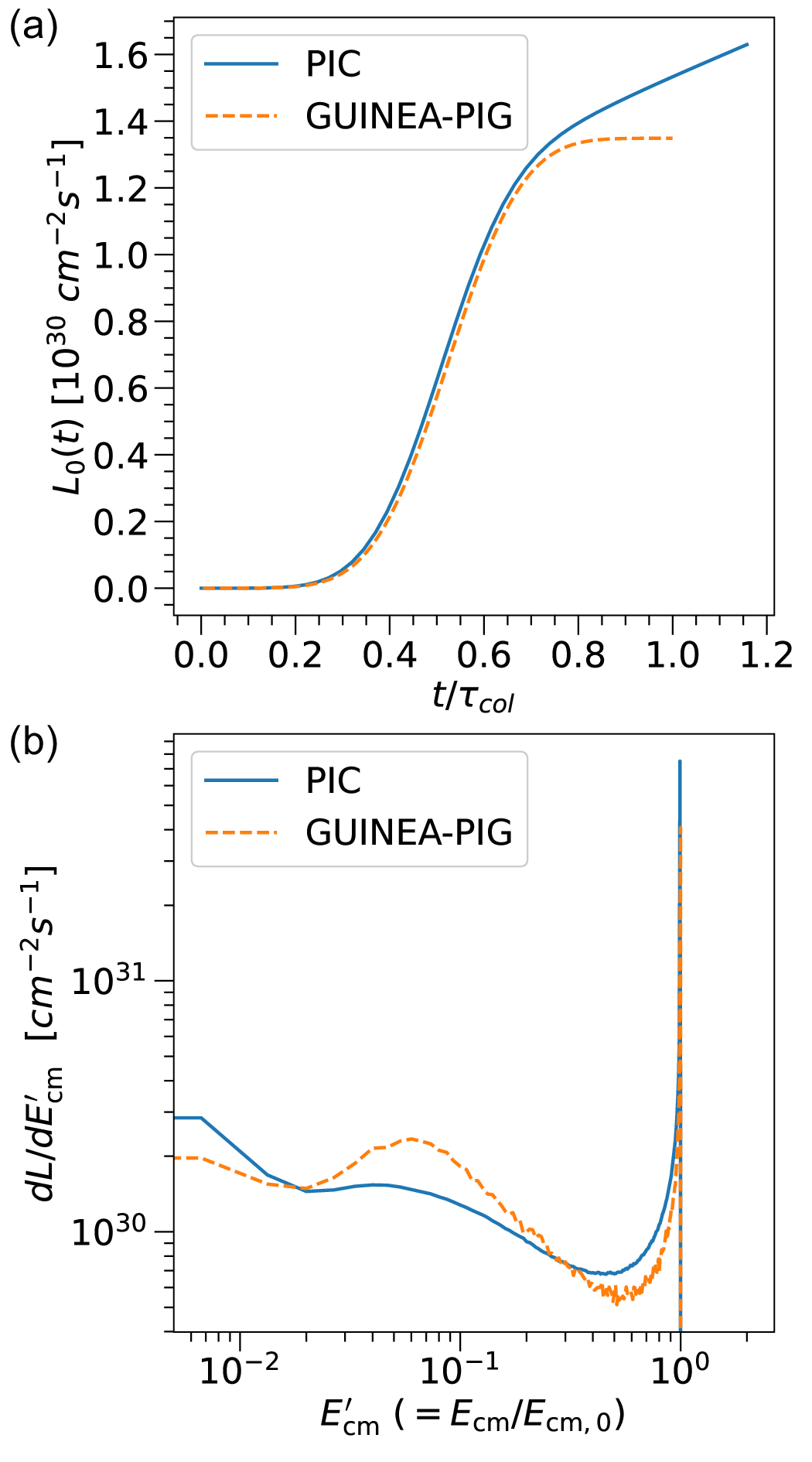}
\caption{(Color online). Luminosity of the collision shown in Fig. \ref{fig: 2D_beams_SFQED_eps_0d2}, obtained by PIC and GUINEA-PIG simulations with SF-QED processes. (a) Growth of the total luminosity $L_0(t)$. (b) Luminosity spectrum $dL/dE_{\mathrm{cm}}'$. $E_{\mathrm{cm}}'=E_{\mathrm{cm}}/E_{\mathrm{cm,0}}$ is the normalized center-of-mass energy, where $E_{\mathrm{cm,0}}=2\mathcal{E}_0$ is the nominal center-of-mass energy.}
\label{fig: lumi_SFQED_eps_0d2}
\end{figure}

\emph{Impact on collision luminosity}---The interplay between SF-QED processes and $\varepsilon$-governed dynamics also dictates the property of collision luminosity. SF-QED processes generally enhance the total luminosity, since the beamstrahlung-induced energy loss leads to increased disruption and beam pinch \cite{Samsonov2021, Zhang2023}. When $\varepsilon$ becomes considerable, the braking effect will elongate the beams, prolong the collision, and allow a longer (more sustainable) duration of beam pinch, as observed in Fig. \ref{fig: 2D_beams_SFQED_eps_0d2}(c). This situation enhances the luminosity, as illustrated in Fig. \ref{fig: lumi_SFQED_eps_0d2}(a). Note that this luminosity enhancement is not captured in the GUINEA-PIG simulations.

The impact on the luminosity spectrum is more nuanced. For technical details on the luminosity spectrum, refer to Appendix \ref{SM_sec: luminosity_spectrum}. Beamstrahlung broadens the luminosity spectrum, which can lead to a potential degradation of collider performance \cite{Chen1992, Poss2014, Schulte2017, Barklow2023, He2025_SLAC_analytical_beamstrahlung}. This broadening effect is also observed in our study. Figure \ref{fig: lumi_SFQED_eps_0d2}(b) illustrates that the luminosity spectrum is dispersed toward lower center-of-mass energies ($E_{\mathrm{cm}}$), primarily driven by the effects of SF-QED.

The PIC luminosity at high $E_{\mathrm{cm}}$ outperforms the GUINEA-PIG result (see Fig. \ref{fig: lumi_SFQED_eps_0d2}(b) for $E_{\mathrm{cm}}/E_{\mathrm{cm,0}} > 0.3$). Conversely, GUINEA-PIG shows larger luminosity at intermediate $E_{\mathrm{cm}}$ values (for $0.02 < E_{\mathrm{cm}}/E_{\mathrm{cm,0}} < 0.3$). These discrepancies arise because GUINEA-PIG overestimates the perpendicular Lorentz force $F_{\mathrm{L, \perp}}$, as it does not account for the braking effect (deceleration of $v_z$). Due to this overestimation, GUINEA-PIG gives an unphysically higher yield of photon emission and pair production. Specifically for the case shown in Fig. \ref{fig: lumi_SFQED_eps_0d2}, the number of coherent pairs produced in GUINEA-PIG is $n_{coh}=0.3$ per primary particle, which is $15 \%$ higher than the PIC result. Additionally, the average energy of a pair particle in GUINEA-PIG is approximately $\mathcal{E}_{p} \simeq 0.12\mathcal{E}_0$, indicating a $21 \%$ increase over the PIC result. The overestimated SF-QED effects result in a reduced luminosity at high $E_{\mathrm{cm}}$ but an increased luminosity at intermediate $E_{\mathrm{cm}}$ in GUINEA-PIG as shown in Fig. \ref{fig: lumi_SFQED_eps_0d2}(b).

In contrast, the PIC luminosity at very low $E_{\mathrm{cm}}$ is higher than that of GUINEA-PIG (see Fig. \ref{fig: lumi_SFQED_eps_0d2}(b) for $E_{\mathrm{cm}}/E_{\mathrm{cm,0}} < 0.02$). As $\varepsilon$ increases due to SF-QED, the particles attain relativistic transverse velocities and experience significant longitudinal deceleration. Consequently, a notable portion of the beams ultimately propagates backward, as illustrated in Fig. \ref{fig: 2D_beams_SFQED_eps_0d2}(c). This substantial redistribution of momentum leads to a marked reduction in $E_{\mathrm{cm}}$ of the corresponding collision events. In particular, $E_{\mathrm{cm}}$ for the reversed particles can approach zero, resulting in greater PIC luminosity at low $E_{\mathrm{cm}}$ compared to GUINEA-PIG. Figure \ref{fig: schematic_SFQED_vs_eps} provides a schematic representation illustrating the causes of low-$E_{\mathrm{cm}}$ collision events.

The luminosity deviations between these two types of simulations are expected to worsen with higher $\varepsilon$ (or $\kappa$).

\emph{Existing beam-beam codes revisited}---The commonly used beam-beam codes, including GUINEA-PIG \cite{Schulte1996} and CAIN \cite{Yokoya1985, Chen1995_ABEL, CAIN244_Manual_2017}, are based on a simplified 2D electrostatic (ES) model. This model is based on the ``free-streaming" assumption where the particles travel at the speed of light ($c$) throughout a collision. This assumption requires that the transverse deflection is negligible ($|v_r| \ll c$). A less stringent requirement is that the number of particles undergoing large-angle deflections is minimal \cite{CAIN244_Manual_2017}. Under the free-streaming simplification, the beams are divided into discrete transverse slices, within which the fields are treated as purely transverse. The fields of each beam (denoted by the subscript $j$) are calculated using the following equations:
\begin{equation}
    \nabla^2 \varphi _j = -4\pi \rho_j, \ E_{r,j} =- \nabla \varphi_j, \ E_{z,j}=0,
    \label{Eq: 2D_ESmodel_fields_GUINEAPIG}
\end{equation}
where $\varphi$ is the electric potential and $\rho$ is the charge density. The magnetic field $B_{\theta,j}$ is determined by Eq. \eqref{Eq: undisrupted_selffields}. With the free-streaming treatment, one has $|B_{\theta,j} | = |E_{r,j}|$. The total system fields are given by
\begin{equation}
    E_r = \sum_j E_{r,j}, \ B_\theta = \sum_j B_{\theta,j} .
    \label{Eq: tot_Er_Btheta_GUINEAPIG}
\end{equation}
In the $\varepsilon \ll 1$ limit, the fully electromagnetic (EM) solver and the 2D ES solver yield very similar results. However, in scenarios with strong deflections or strong SF-QED effects, beam collisions require fully EM simulations, as demonstrated in this study. A 3D ES solver, such as that implemented in WarpX \cite{Vay2008_ESsolver, LBNL_WarpX_collision_2026}, represents a promising intermediate alternative: operating beyond the constraints of the 2D quasi-electrostatic approximation, a 3D ES solver is expected to capture key features, most notably the emergence of the longitudinal field, while remaining less computationally demanding than a fully EM treatment. Some differences with respect to EM results, particularly in the field calculation and its subsequent impact on particle motion, may nonetheless arise for $\varepsilon \ge 1$. Ultimately, for collisions characterized by substantial $\varepsilon$ and $\kappa$, fully self-consistent EM simulation tools remain the most reliable approach for accurate beam-beam modeling.

\emph{Considerations for future high-energy $e^-e^+$ linear colliders}---Our study indicates that the low-energy particles, resulting from quantum beam-beam effects, will be deflected at large angles or even be completely back-scattered (i.e., particle reversal). An example is shown in Fig. \ref{fig: 2D_beams_SFQED_eps_0d2}(c). With the Gaussian-beam geometric factor $\eta=0.15$, we obtain $\varepsilon \simeq 0.0008$ for the ILC $500\ \mathrm{GeV}$ baseline ($N_0=2\times 10^{10}$, $\sigma_z=300\ \mathrm{\mu m}$, $\mathcal{E}_0=250\ \mathrm{GeV}$) \cite{ILC_TDR} and $\varepsilon \simeq 0.0004$ for CLIC at $3\ \mathrm{TeV}$ ($N_0=3.7\times 10^9$, $\sigma_z=44\ \mathrm{\mu m}$, $\mathcal{E}_0=1.5\ \mathrm{TeV}$) \cite{CLIC_updated_baseline}. Both machines operate firmly in the $\varepsilon \ll 1$ regime, where SF-QED effects are weak and $\varepsilon$-governed dynamics are negligible, confirming that codes such as GUINEA-PIG and CAIN remain fully valid for their design and study. It is nonetheless worth noting that CLIC at $3\ \mathrm{TeV}$ can produce a notable number of coherent pairs ($n_{\mathrm{coh}}\sim 0.1$ per beam particle) \cite{Schulte2017, Esberg2014}, which may be susceptible to large-angle deflections or back-scattering. In contrast, proposed $\mathrm{TeV}$-scale plasma-based colliders \cite{Barklow2023, Schroeder2023, Roser2023, 10TeV_design_initiative, He2025_SLAC_analytical_beamstrahlung, Foster2023, Foster2025, Abramowicz2026, Qian2026, ALEGRO2024, ALEGRO2026, Dalichaouch2025, Caldwell2025_ALIVEprogram, Gschwendtner2025_AWAKEinputeuropeanstrategy, zhang2026_TeVebeam, Chigusa2025, Fraser2026} operating with short, high-charge bunches will exhibit notable SF-QED effects \cite{Barklow2023, Zhang2025, He2025_SLAC_analytical_beamstrahlung}. The newly produced pairs, carrying energies several orders of magnitude lower than those of the primary beams, are particularly prone to large $\varepsilon$ and the associated strong $\varepsilon$-governed dynamics (including complete back-scattering) with direct consequences for beam-beam background signals and experimental diagnostics.

\section{Conclusions}
\label{sec: conclusion}
In this study, we introduce a new dimensionless parameter, $\varepsilon$, which dictates the dynamics of beams and fields in high-energy $e^-e^+$ beam collisions. We identify the regime where $\varepsilon \ge 1$ as a threshold for the emergence of new behaviors, notably the dramatic longitudinal deceleration (braking effect) that can lead to complete cessation and reversal of beam propagation. A strong longitudinal electric field is also induced when the beams are perturbed by disruptions. Our theoretical model is consistently validated against electromagnetic particle-in-cell (PIC) simulations, underscoring that prior research has predominantly addressed scenarios where $\varepsilon \ll 1$.

Our findings are summarized as follows:

\begin{enumerate}

\item For $\varepsilon \ll 1$, transverse deflection remains negligible compared to longitudinal streaming, simplifying the particle motion to a harmonic oscillator. However, as $\varepsilon$ exceeds $0.1$, the transverse motion transitions to relativistic speeds, thereby inducing nonlinear dynamics.

\item For finite values of $\varepsilon < 1$, we present an analytical solution indicating that braking effect and energy gain (through transverse field $E_r$) scale with $\varepsilon^2$.

\item Collective dynamics of beams, including density pinch and current generation, are jointly governed by $D$ and $\varepsilon$. During collisions, an axial electric field $E_z$ is produced, scaling as $E_z \simeq \varepsilon E_0$, which results in momentum and energy losses.

\item Analysis of collisions with high $\varepsilon$ ($\varepsilon \ge 1$), conducted through PIC simulations, reveals intricate ring structures and elongated beams, attributed to significant particle reversal resulting from intense deceleration. These features are missing in the GUINEA-PIG simulations.

\item Interplay between SF-QED and dynamics governed by $\varepsilon$ is characterized by another parameter $\kappa$ introduced in this study. For $\kappa \gtrsim 0.1$, the $\varepsilon$-governed dynamics become pronounced, potentially leading to beam and current reversal and restricting quantum processes.

\item Luminosity exhibits a rapid increase in the high-$\varepsilon$ regime, primarily due to extended collision durations and pronounced and more sustainable beam pinch. Furthermore, the influence of $\varepsilon$ on the luminosity spectrum demonstrates that while it enables a high luminosity at high center-of-mass energies $E_{\mathrm{cm}}$ by mitigating SF-QED effects, it concurrently results in collision events characterized by very low $E_{\mathrm{cm}}$ due to a severe braking effect.

\end{enumerate}

Finally, our study highlights the necessity of using more self-consistent simulation tools for studying collisions characterized by substantial $\varepsilon$ (or $\kappa$). This finding may be pertinent to the design and operation of future $\mathrm{TeV}$-class plasma-based linear colliders \cite{Barklow2023, Schroeder2023, Roser2023, 10TeV_design_initiative, He2025_SLAC_analytical_beamstrahlung, Foster2023, Foster2025, Abramowicz2026, Qian2026, ALEGRO2024, ALEGRO2026, Dalichaouch2025, Caldwell2025_ALIVEprogram, Gschwendtner2025_AWAKEinputeuropeanstrategy, zhang2026_TeVebeam, Chigusa2025, Fraser2026}, where notable SF-QED effects are anticipated. In such scenarios, $\varepsilon$ is elevated due to SF-QED phenomena. The produced secondary particles can have considerable $\varepsilon$ due to their very low energies and may undergo large-angle deflection or complete back-scattering. Therefore, the implementation of more self-consistent numerical tools—especially those that accurately model fields and particle motion—is essential for reliable beam-beam simulations, such as PIC codes or other advanced solvers \cite{Vay2008_ESsolver, LBNL_WarpX_collision_2026}.

\begin{acknowledgments}
We acknowledge Professor Warren Mori (UCLA) for the inspiring discussions. W.Z. acknowledges Yuxue Zhang (China Academy of Engineering Physics) for the kind assistance with PARATERA HPC. W.Z. is supported by Jiangxi Provincial Natural Science Foundation (the Young Scientists Fund, No. 20242BAB21003), Start-up Fund (No. DHBK2023009) from East China University of Technology (ECUT), and the East China Accelerator $\&$ Neutron Source (ECANS) project at ECUT. This work is also supported by FCT (Portugal) Grants No. 2022.02230.PTDC (X-MASER), UIDB/FIS/50010/2020 - PESTB 2020-23, Grants No. CEECIND/04050/2021 and PTDC/FIS-PLA/ 3800/2021. We acknowledge Beijng PARATERA Tech CO., Ltd. for providing HPC resources. Simulations were performed at National Supercomputer Center (Guangzhou) and PARATERA HPC (Beijing).   
\end{acknowledgments}

\appendix

\section{Perturbation solution to the particle motion for finite $\varepsilon$}
\label{SM_sec: perturbation_solution}

We solve the particle motion in a perturbation manner for finite $\varepsilon$ (still satisfying $\varepsilon < 1$). The equation of motion is given by Eq. \eqref{Eq: equation_motion_beta_gamma} where an electron under study is driven by Lorentz force. The Lorentz force is given by $\boldsymbol{F_\mathrm{L}}=-e(\boldsymbol{E_r}+\boldsymbol{v}\times \boldsymbol{B_\theta}/c)$, where $\boldsymbol{v}$ is the particle velocity, $\boldsymbol{E_r}$ and $\boldsymbol{B_\theta}$ are the fields of the opposing (positron) beam. The longitudinal field $E_z$ is not considered here.

Equation \eqref{Eq: equation_motion_beta_gamma} can be rearranged as:
\begin{subequations}
\begin{align}
\gamma \frac{d^2r}{dt^2} & = -\frac{1}{2}(1+\beta_z)\omega_b^2r + \frac{1}{2}\omega_b^2\beta_r^2r, \label{Eq: equation_r_forperturbation} \\
\frac{dp_z}{dt} & = \frac{1}{2}\varepsilon\beta_r\frac{r}{\sigma_0}\frac{\gamma_0mc}{\tau_D}, \label{Eq: equation_pz_forperturbation} \\
\frac{d\gamma}{dt} & = -\frac{1}{2}\varepsilon\beta_r\frac{r}{\sigma_0} \frac{\gamma_0}{\tau_D}, \label{Eq: equation_gamma_forperturbation} \\
\frac{d\beta_z}{dt} & = \frac{1}{2}(1+\beta_z)\beta_r \frac{\gamma_0}{\gamma}\frac{r}{\sigma_0}\frac{\varepsilon}{\tau_D}. \label{Eq: equation_betaz_forperturbation}
\end{align}
\label{Eq: equation_motion_forperturbation}
\end{subequations}

We look for a perturbative solution expanded as a power series in $\varepsilon$, i.e.,
\begin{subequations}
\begin{align}
r & = r^{(0)} + \varepsilon r^{(1)} + \varepsilon^2 r^{(2)} + ..., \label{Eq: r_perturbation_expression} \\
\Omega & = \Omega^{(0)} + \varepsilon \Omega^{(1)} + \varepsilon^2 \Omega^{(2)} + ..., \label{Eq: Omega_perturbation_expression} \\
\gamma & = \gamma^{(0)} + \varepsilon \gamma^{(1)} + \varepsilon^2 \gamma^{(2)} + ..., \label{Eq: gamma_perturbation_expression} \\
p_z & = p_z^{(0)} + \varepsilon p_z^{(1)} + \varepsilon^2 p_z^{(2)} + ... , \label{Eq: pz_perturbation_expression}
\end{align}
\label{Eq: quantities_perturbation_expression}
\end{subequations}
where the superscript $(n)$ represents the $n\mathrm{th}$-order quantities. $\Omega$ is the frequency of nonlinear oscillatory motion of the particle. 

At the $0\mathrm{th}$ order, the relevant quantities are given by
\begin{subequations}
\begin{align}
& \Omega^{(0)} = \frac{1}{\tau_D}=\frac{\omega_b}{\sqrt{\gamma_0}}, \ \gamma^{(0)}=\gamma_0, \label{Eq: 0order_quantities_1} \\
& p_z^{(0)}=p_{z,0} \simeq \gamma_0mc, \ \beta_{z}^{(0)}=\beta_{z,0}\simeq 1, \label{Eq: 0order_quantities_2}
\end{align}
\label{Eq: 0order_quantities}
\end{subequations}
where $\Omega^{(0)}$ is the fundamental frequency for the ideal harmonic oscillator (Sec. \ref{subsec: harmonic_oscillator}). The $0\mathrm{th}$-order particle motion is given by
\begin{subequations}
\begin{align}
& r^{(0)} = r_0\cos(\Omega t), \label{Eq: 0order_r} \\
& \beta_r^{(0)} = - \frac{r_0\Omega}{c}\sin(\Omega t) = -\varepsilon \frac{r_0}{\sigma_0}\frac{\Omega}{\Omega^{(0)}}\sin(\Omega t). \label{Eq: 0order_betar}
\end{align}
\label{Eq: 0order_r_betar}
\end{subequations}
Since $\beta_r^{(0)}\sim \mathcal{O}(\varepsilon)$, it is straightforward to infer using Eq. \eqref{Eq: equation_motion_forperturbation} that all the $1\mathrm{st}$-order quantities are vanishing. By injecting the $0\mathrm{th}$-order solution [Eq. \eqref{Eq: 0order_r_betar}] into Eq. \eqref{Eq: equation_motion_forperturbation}, we obtain
\begin{subequations}
\begin{align}
\gamma^{(2)} & = \frac{1}{4}\frac{r_0^2}{\sigma_0^2}\left(1-\cos^2 (\Omega^{(0)} t)\right) \gamma_0, \label{Eq: gamma_2nd_order} \\
p_z^{(2)} & = -\frac{1}{4}\frac{r_0^2}{\sigma_0^2}\left(1-\cos^2 (\Omega^{(0)} t)\right) \gamma_0 mc, \label{Eq: pz_2nd_order} \\
\beta_z^{(2)} & = -\frac{1}{2}\frac{r_0^2}{\sigma_0^2}\left(1-\cos^2 (\Omega^{(0)} t)\right). \label{Eq: betaz_2nd_order}
\end{align}
\label{Eq: quantities_perturbation_2nd_order}
\end{subequations}
Therefore, the perturbation solution under the Lorentz force $F_\mathrm{L}$ (without considering $E_z$) is given by
\begin{subequations}
\begin{align}
\gamma ^{F_\mathrm{L}} & =\gamma_0\left[1 + \frac{1}{4}\varepsilon^2\frac{r_0^2}{\sigma_0^2}\left(1-\cos^2 (t/\tau_D) \right) \right], \label{Eq: SM_perturbation_solution_gamma} \\
p_z ^{F_\mathrm{L}}& = p_{z,0} \left[1 -\frac{1}{4}\varepsilon^2\frac{r_0^2}{\sigma_0^2}\left(1-\cos^2 (t/\tau_D)\right) \right], \label{Eq: SM_perturbation_solution_pz} \\
\beta_z ^{F_\mathrm{L}} & = \beta_{z,0} \left[1 -\frac{1}{2}\varepsilon^2\frac{r_0^2}{\sigma_0^2}\left(1-\cos^2 (t/\tau_D)\right) \right]. \label{Eq: SM_perturbation_solution_betaz}
\end{align}
\label{Eq: SM_perturbation_solution}
\end{subequations}
In the early interaction phase ($t\ll \tau_D$), Eq. \eqref{Eq: SM_perturbation_solution} gives
\begin{subequations}
\begin{align}
    \frac{\Delta \gamma ^{F_\mathrm{L}}}{\gamma_0} &\sim \frac{1}{4}\varepsilon^2\frac{r_0^2}{\sigma_0^2}\frac{t^2}{\tau_D^2}, \label{Eq: SM_early_perturbation_gamma} \\
    \frac{\Delta p_z ^{F_\mathrm{L}}}{p_{z,0}} &\sim -\frac{1}{4}\varepsilon^2\frac{r_0^2}{\sigma_0^2}\frac{t^2}{\tau_D^2}, \label{Eq: SM_early_perturbation_pz} \\
    \frac{\Delta \beta_z ^{F_\mathrm{L}}}{\beta_{z,0}} &\sim -\frac{1}{2}\varepsilon^2\frac{r_0^2}{\sigma_0^2}\frac{t^2}{\tau_D^2}. \label{Eq: SM_early_perturbation_betaz}
\end{align}
\label{Eq: SM_early_perturbation}
\end{subequations}

\section{$E_z$-caused momentum and energy loss}
\label{SM_sec: Deltapz_Deltagamma_by_Ez}
A non-zero longitudinal field, $E_z$, is generated by the disruption process during a collision, as derived in Sec. \ref{subsec: field_Ez}. This field results in momentum and energy losses. 

In the small-$\varepsilon$ regime ($\varepsilon < 1$), we retain only the leading term [$\mathcal{O}(\varepsilon)$] in the formula of $E_z$ [Eqs. \eqref{Eq: full_delta_Ez} and \eqref{Eq: delta_Ez0}] and omit higher-order terms [$\mathcal{O}(\varepsilon^2)$]. With this simplification, the $E_z$-caused momentum loss is given by
\begin{align}
     \Delta p_z ^{E_z} & = \int_0^{t} (-e\delta E_z)dt' \nonumber \\
     & \simeq \int_0^{t} \left(-eE_0\sqrt{D}\varepsilon \frac{t'}{\tau_{col}}\right) \left( 1-\frac{r^2}{\sigma_0^2}\right) dt' \nonumber \\
     & = -\frac{1}{4}\varepsilon^2 \Bigg[\left(1-\frac{1}{2}\frac{r_0^2}{\sigma_0^2}\right)\frac{t^2}{\tau_D^2} \label{Eq: SM_Delta_pz_Ez} \\
     & \ \ \ + \frac{1}{4}\frac{r_0^2}{\sigma_0^2} \left(1 - \cos\frac{2t}{\tau_D} - \frac{2t}{\tau_D}\sin \frac{2t}{\tau_D}\right) \Bigg]p_{z,0}, \nonumber
    \end{align}
Neglecting the braking effect (since $\Delta \beta_z \sim -\varepsilon^2/2 \ll 1$), the energy loss caused by $E_z$ can be expressed as
\begin{equation}
    \Delta \gamma ^{E_z} \simeq \frac{1}{mc^2} \int_0^{t} (-e\delta E_zv_{z,0})dt' = \frac{\Delta p_z ^{E_z}}{mc}.
    \label{Eq: SM_Delta_gamma_Ez}
\end{equation}
Equation \eqref{Eq: SM_Delta_gamma_Ez} can be recast in a convenient form as
\begin{equation}
    \Delta \gamma ^{E_z} \simeq  \frac{\Delta p_z ^{E_z}}{p_{z,0}}\gamma_0.
    \label{Eq: SM_Delta_gamma_gamma0_Ez}
\end{equation}

\section{Particle-in-cell simulations for validating the theoretical model}
\label{SM_sec: PIC_simulation_verify_model}

The theoretical model established in Sec. \ref{sec: model} is validated through three-dimensional (3D), fully electromagnetic, particle-in-cell (PIC) simulations conducted using OSIRIS. 

Our simulations employ the same beams considered by the theoretical model (see Fig. \ref{fig: schematic_collision_setup_model} as a schematic plot). We perform a series of simulations with fixed energy of $\mathcal{E}_0 = 10\ \mathrm{GeV}$ and disruption parameter of $D = 1.88$. The profile ratio $\sigma_0/\sigma_z$ is varied to achieve different values of $\varepsilon$ according to its definition [Eq. \eqref{Eq: eps_definition}]. This results in a broad range of $\varepsilon$ values ($0.01 \leq \varepsilon \leq 1.7$), with detailed beam parameters summarized as follows:
\begin{enumerate}
    \item $\varepsilon = 0.01$: $\sigma_0=4.44\ \mathrm{nm}$, $\sigma_z=565.7\ \mathrm{nm}$, $N_0=1.14\times 10^{8}$, $n_0=3.25\times 10^{24}\ \mathrm{cm^{-3}}$
    \item $\varepsilon = 0.043$: $\sigma_0=8.88\ \mathrm{nm}$, $\sigma_z=282.85\ \mathrm{nm}$, $N_0=9.1\times 10^{8}$, $n_0=1.3\times 10^{25}\ \mathrm{cm^{-3}}$
    \item $\varepsilon = 0.067$: $\sigma_0=11.1\ \mathrm{nm}$, $\sigma_z=226.28\ \mathrm{nm}$, $N_0=1.78\times 10^{9}$, $n_0=2.03\times 10^{25}\ \mathrm{cm^{-3}}$
    \item $\varepsilon = 0.12$: $\sigma_0=14.8\ \mathrm{nm}$, $\sigma_z=169.71\ \mathrm{nm}$, $N_0=4.21\times 10^{9}$, $n_0=3.61\times 10^{25}\ \mathrm{cm^{-3}}$
    \item $\varepsilon = 0.27$: $\sigma_0=22.2\ \mathrm{nm}$, $\sigma_z=113.14\ \mathrm{nm}$, $N_0=1.42\times 10^{10}$, $n_0=8.12\times 10^{25}\ \mathrm{cm^{-3}}$
    \item $\varepsilon = 0.48$: $\sigma_0=29.6\ \mathrm{nm}$, $\sigma_z=84.86\ \mathrm{nm}$, $N_0=3.36\times 10^{10}$, $n_0=1.44\times 10^{26}\ \mathrm{cm^{-3}}$
    \item $\varepsilon = 1.08$: $\sigma_0=44.4\ \mathrm{nm}$, $\sigma_z=56.57\ \mathrm{nm}$, $N_0=1.14\times 10^{11}$, $n_0=3.25\times 10^{26}\ \mathrm{cm^{-3}}$
    \item $\varepsilon = 1.7$: $\sigma_0=55.72\ \mathrm{nm}$, $\sigma_z=45.07\ \mathrm{nm}$, $N_0=2.25\times 10^{11}$, $n_0=5.11\times 10^{26}\ \mathrm{cm^{-3}}$
\end{enumerate}
The simulations described above yield the results presented in Fig. \ref{fig: current_field_unif_beam}, Fig. \ref{fig: phase_Ez_unif_beam}, and Fig. \ref{fig: delta_pz_eps_unif_beam}.

These simulations are conducted in Cartesian coordinates $(x, y, s)$, with the electron and positron beams propagating in $+s$ and $-s$ directions, respectively. Here, we provide details for the specific case where $\varepsilon = 0.12$. The simulation domain is defined as $-10\sigma_0 \leq x \leq 10\sigma_0$, $-10\sigma_0 \leq y \leq 10\sigma_0$, and $-\sigma_z \leq s \leq \sigma_z$, divided into a grid of $500 \times 500 \times 840$ cells. This configuration results in spatial resolutions of $\Delta x = \Delta y = 0.04\sigma_0$ and $\Delta s = 0.0024\sigma_z$. The time step is set to $\Delta t = 0.001\sigma_z/c$.

In the simulations, the electron beam is initially positioned in the region $-\sigma_z \leq s \leq 0$, while the positron beam occupies $0 \leq s \leq \sigma_z$. They are modeled as cold and initialized with $8$ macro-particles per cell, yielding a total of $6.54 \times 10^6$ macro-particles for each beam. At the front and tail of the beams, the density rapidly decreases to zero within a narrow layer of width $d_0 = 0.052\sigma_z$, implemented using a sine function. The simulation runs until $t = \tau_{col} = \sigma_z/c$, marking the nominal end of the collision.

Other collision cases utilize similar setups. For high-$\varepsilon$ cases, an extended simulation duration (beyond $\tau_{col}$) and larger simulation domains in the $s$ direction are necessary. This requirement arises from the braking effect; with high $\varepsilon$, a significant fraction of particles decelerate in $v_z$ and later reverse their propagation. Consequently, the beams become elongated, resulting in prolonged collisions.

\section{Simulations for high-$\varepsilon$ collisions}
\label{SM_sec: PIC_GP_simulation_high_eps}

Both PIC (using OSIRIS) and GUINEA-PIG simulations are conducted to investigate high-$\varepsilon$ collisions, as discussed in Sec. \ref{sec: high_eps_simulations}. We have specifically examined a case with $\varepsilon = 3.6$, where SF-QED effects, including photon emission (beamstrahlung) and electron-positron pair production, are disabled in the simulations. The corresponding results are presented in Fig. \ref{fig: 3D_beam_fourmoments_eps_2d5} and Fig. \ref{fig: 2D_beamslice_eps_2d5}.

The beam profiles are Gaussian, characterized by the density $n = n_0 \exp\left(-\frac{x^2 + y^2}{2\sigma_0^2}\right) \exp\left(-\frac{(s - s_0)^2}{2\sigma_z^2}\right)$, where $\sigma_0$ is the root mean square (RMS) transverse size, $\sigma_z$ is the RMS longitudinal length, and $s_0$ is the beam center. For the case with $\varepsilon = 3.6$, the parameters are: $\mathcal{E}_0 = 10\ \mathrm{GeV}$, $N_0 = 1.12 \times 10^{12}$, $\sigma_z = 20\ \mathrm{nm}$, $\sigma_0 = 20\ \mathrm{nm}$, and $n_0 = 8.9 \times 10^{27}\ \mathrm{cm}^{-3}$. A finite emittance of $\epsilon = 100\ \mathrm{nm}$ is implemented for both beams.

The OSIRIS simulation employs a computational box of $18\sigma_0 \times 18\sigma_0 \times 12\sigma_z$, divided into $480 \times 480 \times 2400$ grids, resulting in resolutions of $\Delta x = \Delta y = 0.037\sigma_0$ and $\Delta z = 0.005\sigma_z$. The time step is set to $\Delta t = 0.004\sigma_z/c$. The simulation utilizes $4.8 \times 10^7$ macro-particles for each beam, corresponding to $2$ macro-particles per cell. Initially, the electron and positron beams are centered at $s_0 = -3\sigma_z$ and $3\sigma_z$, respectively. The beams are truncated for $\sqrt{x^2 + y^2} > 3\sigma_0$ and for $|s - s_0| > 3\sigma_z$. Emittance is introduced by assigning thermal velocities to the particles in the transverse direction.

\begin{figure}
\includegraphics[width=5.5cm,height=13.9cm]{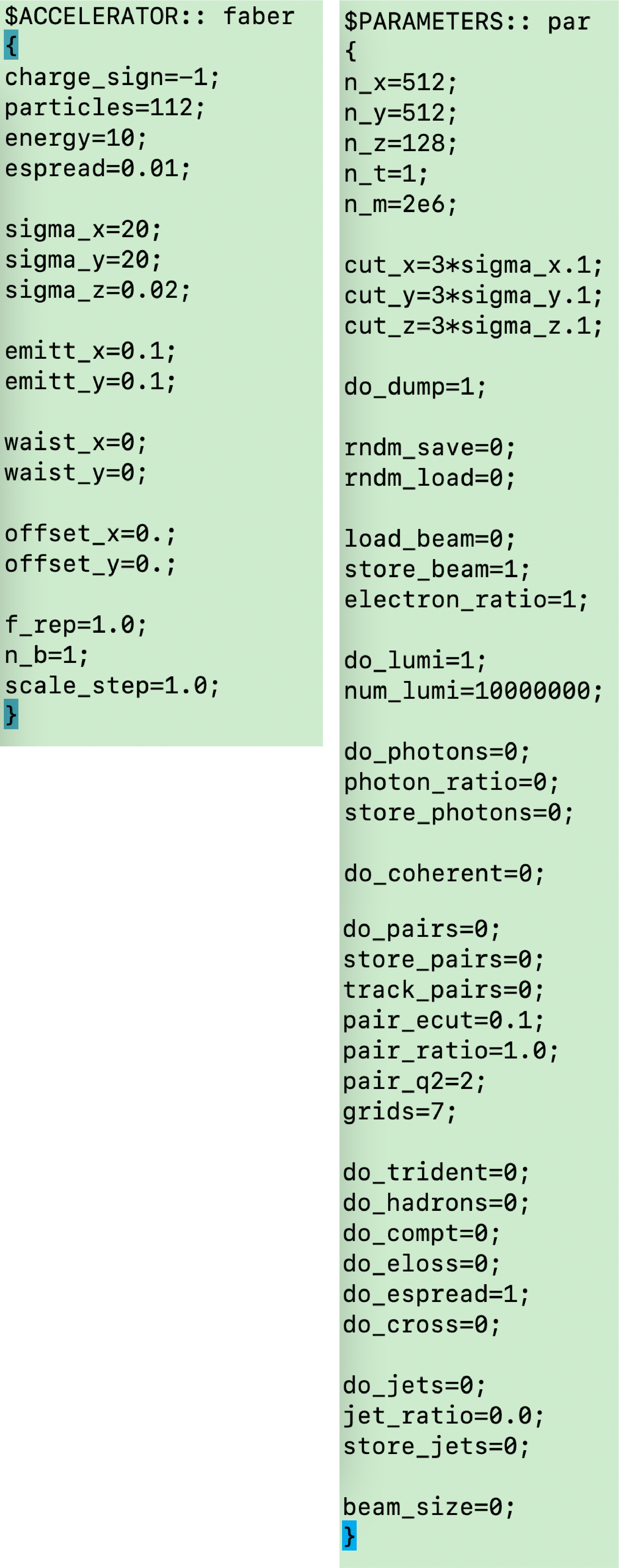}
\caption{(Color online). Snapshot of the input file of the GUINEA-PIG simulation for exploring high-$\varepsilon$ collisions (Sec. \ref{sec: high_eps_simulations} and Appendix \ref{SM_sec: PIC_GP_simulation_high_eps}). SF-QED is disabled here.}
\label{fig: GP_input_eps_2d5}
\end{figure}

The GUINEA-PIG simulation is conducted using guineapig++ (Version 1.2.2). The setup for GUINEA-PIG matches that of the OSIRIS simulation. A snapshot of the input file for GUINEA-PIG simulation is provided in Fig. \ref{fig: GP_input_eps_2d5}. Each beam is initialized with $2 \times 10^6$ macro-particles. The number of substeps is set to $n_t = 1$. The transverse resolution is configured as $n_x = n_y = 512$, with the beams divided into $n_z = 128$ slices.

\section{Distinction between SF-QED processes and $\varepsilon$-governed dynamics}
\label{SM_sec: distinction_SFQED_epsdynamics}

Beamstrahlung predominantly affects the outer particles, particularly those around the periphery, located at ($r \sim \sigma_0$), where the fields are most intense. In contrast, the $\varepsilon$-governed dynamics apply to the entire region, as discussed in Sec. \ref{subsec: overall_changes_pz_gamma}. Specifically, the induced field $E_z$ in an electron-positron collision reaches its maximum at the axis. Therefore, the operational regions for beamstrahlung and $\varepsilon$-governed dynamics do not fully overlap. Furthermore, beamstrahlung alone cannot completely arrest the particles or even reverse their propagation, since the quantum parameter (characterizing the quantum strength \cite{Zhang2023}) decreases with energy loss (decreasing $\gamma$). These features highlight the distinction between SF-QED and the $\varepsilon$-governed dynamics.

\section{Simulations for the impact of SF-QED}
\label{SM_sec: PIC_GP_SFQED_simulation}

\begin{figure}
\includegraphics[width=5.4cm,height=13.9cm]{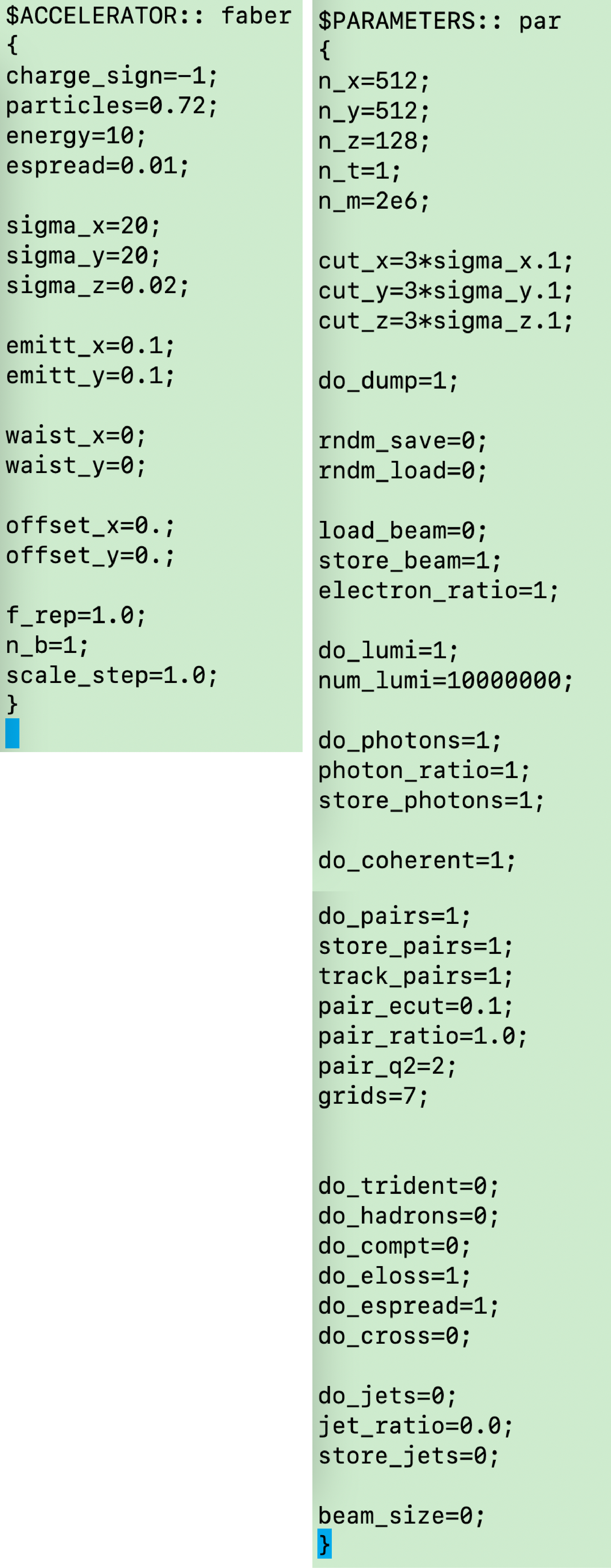}
\caption{(Color online). Snapshot of the input file of the GUINEA-PIG simulation for studying the impact of SF-QED (Sec. \ref{sec: discussion} and Appendix \ref{SM_sec: PIC_GP_SFQED_simulation}).}
\label{fig: GP_input_SFQED_eps_0d2}
\end{figure}

Section \ref{sec: discussion} presented a case study demonstrating the impact of SF-QED on the $\varepsilon$-governed dynamics. The results are presented in Fig. \ref{fig: 2D_beams_SFQED_eps_0d2}. The PIC simulations with OSIRIS have included the beamstrahlung (photon emission) and coherent electron-positron pair production. See \cite{Zhang2025} for details of SF-QED implementation in OSIRIS.

We employed the same beams as those used in Appendix \ref{SM_sec: PIC_GP_simulation_high_eps}, but with a reduced particle number of $N_0 = 7.2 \times 10^9$. The other parameters remained the same: $\mathcal{E}_0 = 10\ \mathrm{GeV}$, $\sigma_0 = 20\ \mathrm{nm}$, $\sigma_z = 20\ \mathrm{nm}$, and emittance $\epsilon = 100\ \mathrm{nm}$. This results in $\varepsilon = 0.28$ and $D = 0.05$. SF-QED effects, including beamstrahlung and coherent electron-positron pair production, are activated.

For the OSIRIS simulation, a computational box of $20\sigma_0 \times 20\sigma_0 \times 14\sigma_z$ is utilized, divided into $360 \times 360 \times 1800$ grids, yielding spatial resolutions of $\Delta x = \Delta y = 0.055\sigma_0$ and $\Delta z = 0.0078\sigma_z$. The time step is set to $\Delta t = 0.007\sigma_z/c$. The simulation employs $2.8 \times 10^7$ macro-particles for each beam, corresponding to $4$ macro-particles per cell. The beam profiles, centers, and truncations are the same with those of the beams in Appendix \ref{SM_sec: PIC_GP_simulation_high_eps}. Initially, the electron and positron beams occupy the regions of $-6\sigma_z\le s \le 0$ and $0 \le s \le 6 \sigma_z$, respectively. To facilitate insight into the late interaction stage, a longer simulation box along the $s$ axis is used here, and the simulation runs until the beam fronts reach the boundary of the box at $t = 7\sigma_z/c$.

The GUINEA-PIG simulation closely resembles the one performed in Appendix \ref{SM_sec: PIC_GP_simulation_high_eps}. Figure \ref{fig: GP_input_SFQED_eps_0d2} displays the input file for this GUINEA-PIG simulation. The GUINEA-PIG results indicate that coherent pair production is approximately $10^{5}$ times higher than incoherent pair production, demonstrating the significant effects of SF-QED in this scenario.

\section{Luminosity spectrum and simulation data processing}
\label{SM_sec: luminosity_spectrum}

The center-of-mass energy of a collision between two particles is given by
\begin{equation}
   E_{\mathrm{cm}} = \sqrt{ (\mathcal{E}_1+\mathcal{E}_2)^2 -c^2(\boldsymbol{p}_1+\boldsymbol{p}_2)^2},
\end{equation}
where $\mathcal{E}$ and $\boldsymbol{p}$ are the energy and (vector) momentum, respectively. 

Luminosity spectrum in a beam-beam collision evaluates the luminosity distribution with respect to $E_{\mathrm{cm}}$. It is defined as
\begin{equation} 
\begin{split}
 & \frac{dL}{dE_{\mathrm{cm}}}(E_{\mathrm{cm}}, t) \\
 & = \int_0^t dt'\int d\boldsymbol{x}\,d\boldsymbol{p}_1 d\boldsymbol{p}_2 
\sqrt{ |\boldsymbol{v}_1 - \boldsymbol{v}_2|^2 - |\boldsymbol{v}_1\times\boldsymbol{v}_2|^2/c^2} \\ 
& \ \ \ f_1(\boldsymbol{x}, \boldsymbol{p}_1, t')f_2(\boldsymbol{x}, \boldsymbol{p}_2, t') \delta(E_{\mathrm{cm}} - E_{\mathrm{cm}}(\boldsymbol{p}_1, \boldsymbol{p}_2)),
\end{split}
\label{Eq: SM_spec_lumi}
\end{equation}
where $\boldsymbol{v}$ is the particle velocity and $f(\boldsymbol{x}, \boldsymbol{p}, t)$ is the particle distribution function.

In this study, the luminosity spectrum from PIC simulations is obtained by analyzing collisions between macro-particles from the two beams within each grid cell.

For the GUINEA-PIG simulation, the luminosity spectrum is computed by examining collision events recorded in the output file ``lumi.ee.out." In GUINEA-PIG, particles are assumed to travel at the speed of light $c$, and their transverse momenta are neglected. Consequently, $E_{\mathrm{cm}}$ in GUINEA-PIG is calculated simply as $E_{\mathrm{cm}} = \sqrt{4\mathcal{E}_1 \mathcal{E}_2}$.

\bibliography{refs}

\end{document}